\documentclass{optica-article}

\journal{opticajournal} 

\articletype{Research Article}

\usepackage{lineno}

\usepackage{graphicx}
\usepackage{dcolumn}
\usepackage{bm}
\usepackage{hyperref}
\hypersetup{
	colorlinks   = true, 
	urlcolor     = blue, 
	linkcolor    = blue, 
	citecolor    = blue 
}
\usepackage{pifont}
\usepackage{multirow}
\usepackage{tabularray}
\usepackage{array}
    \newcolumntype{P}[1]{>{\centering\arraybackslash}p{#1}}
    \newcolumntype{M}[1]{>{\centering\arraybackslash}m{#1}}
\usepackage{stackengine}
    
\usepackage{romannum}
\usepackage{bigints}
\usepackage{ulem}
\usepackage{xcolor}
\usepackage{tabularx} 
\usepackage{siunitx} 
\usepackage[footnotehyper]{nicematrix}
\NiceMatrixOptions{cell-space-limits = 1.4pt}
\usepackage[version=4]{mhchem}
\usepackage{gensymb}

\begin{document}

\title{Beyond the Purcell Effect: Controlling Pure Quantum Dephasing with Spin Noise Metasurfaces}

\author{Wenbo Sun,\authormark{1,2,$\dagger$} Shoaib Mahmud,\authormark{1,2,$\dagger$} Wei Zhang,\authormark{1,2,$\dagger$} Runwei
Zhou,\authormark{1} Pronoy Das,\authormark{1,2} Dan Jiao,\authormark{1} Zubin Jacob,\authormark{1,2,*}}

\address{\authormark{1}Elmore Family School of Electrical and Computer Engineering, Purdue University, West Lafayette, Indiana 47907, USA\\
\authormark{2}Birck Nanotechnology Center, Purdue University, West Lafayette, Indiana 47907, USA\\
\authormark{$\dagger$}The authors contributed equally to this work}

\email{\authormark{*}zjacob@purdue.edu} 


\begin{abstract*} 
One central theme in quantum photonics is tailoring the interactions between atoms/spins and their electromagnetic (EM) environments. Considerable effort has focused on engineering spontaneous emission by shaping EM environments, known as the Purcell effect. However, photonic environment control of pure dephasing, which is a complementary paradigm of non-unitary atom/spin couplings with EM environments, remains largely unexplored. Here, we introduce a nanophotonic approach to modify qubit pure dephasing dynamics. Unlike Purcell engineering that tailors photonic environments at qubit resonance frequencies (typically optical/near-infrared), we develop ultra-subwavelength spin noise metasurfaces for efficient broadband control of low-frequency (e.g., $\sim$MHz) photonic environments far off-resonant with atoms/spins for dephasing engineering. We experimentally demonstrate our approach using lithographically defined CoFeB metasurfaces and shallow nitrogen-vacancy (NV) centers in diamond. Instead of modified spontaneous emission, we observe modified NV pure dephasing dynamics near different spin noise metasurfaces. We further isolate metasurface-controlled dephasing from other dephasing mechanisms (e.g., spin bath) by measuring the NV ensemble dephasing noise spectrum with dynamical decoupling spectral decomposition techniques. Our results establish a new frontier in engineering quantum light-matter interactions with nanophotonic structures.
\end{abstract*}

\section{Introduction}
Pure dephasing and spontaneous emission are two paradigms of non-unitary processes arising from interactions between atoms/spins with their electromagnetic (EM) environments. Spontaneous emission describes the radiative decay of excited states with photon emission. In contrast, pure dephasing refers to the loss of phase coherence in superposition states without energy decay, and is often a main obstacle in quantum technologies. One central theme in quantum photonics is to engineer EM environments and tailor their interactions with atoms/spins for fundamental phenomena and practical applications~\cite{gonzalez2024light,hammerer2010quantum,lodahl2015interfacing,RevModPhys.90.031002,solntsev2021metasurfaces}. Considerable effort has focused on engineering spontaneous emission processes using photonic structures, known as the Purcell effect~\cite{purcell1946resonance}. Meanwhile, controlling pure quantum dephasing dynamics via structured photonic environments remains much less explored.

\begin{figure}[!t]
    \centering
    \includegraphics[width=3.2 in]{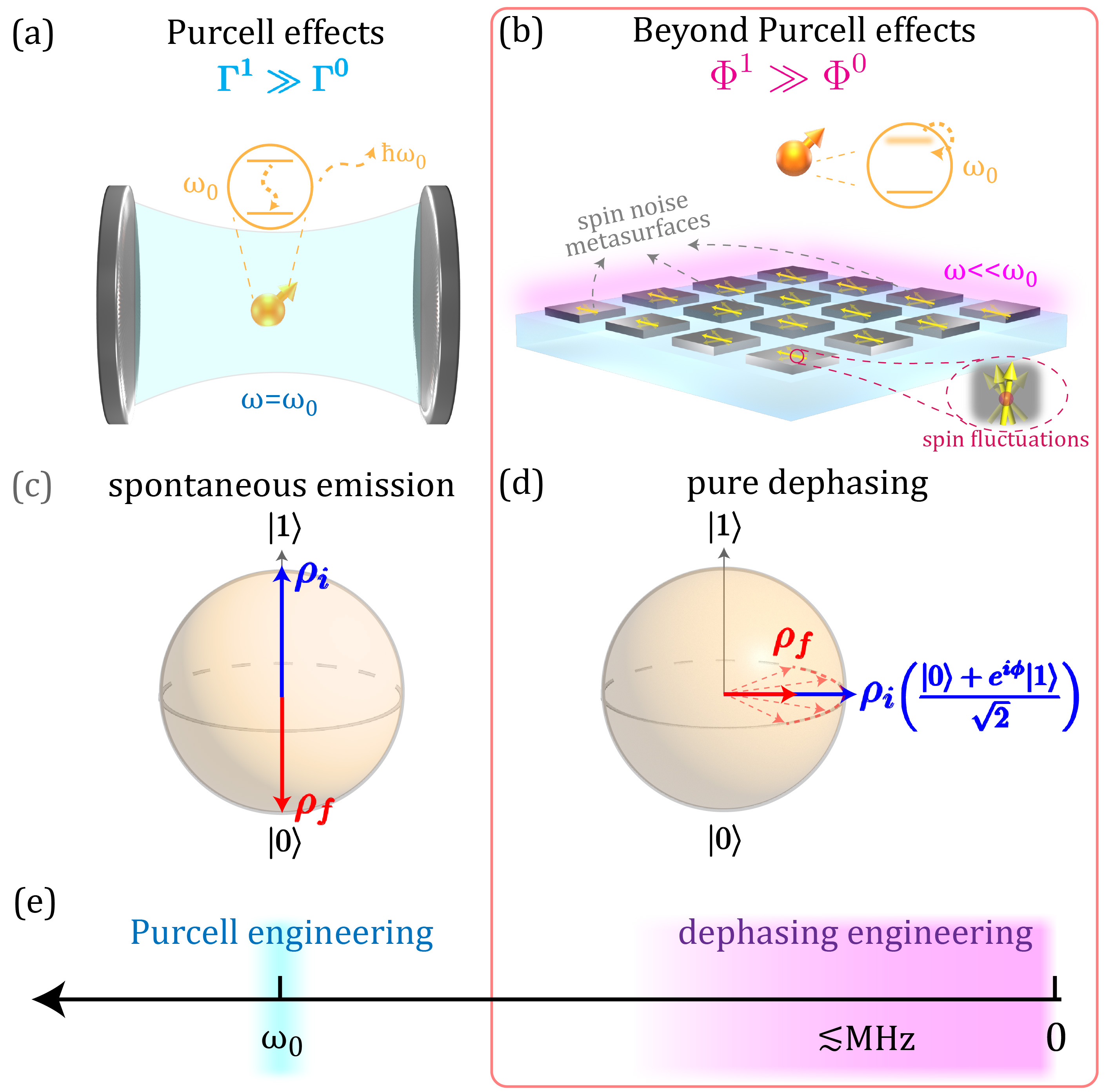}
    \caption{Tailoring pure quantum dephasing with structured nanophotonic environments. (a) The Purcell effect manipulates spontaneous emission with cavity EM environments. (b) In contrast, we propose to control pure dephasing with ultra-subwavelength spin noise metasurfaces. (c) Spontaneous emission describes the radiative decay of excited states $|1\rangle$. (d) Pure dephasing refers to the loss of phase coherence in superposition states $\frac{1}{\sqrt{2}}(|0\rangle + |1\rangle)$ without photon emission. (e) Spontaneous emission is governed by photonic LDOS at atom resonance frequencies $\omega \approx \omega_0$, while pure dephasing is determined by broadband low-frequency spectral components $\omega \ll \omega_0$ of the EM environments.}
    \label{fig:fig1}
\end{figure}

Spontaneous emission is governed by photonic local density of states (LDOS) at atom resonance frequencies (typically optical/near-infrared) [see Fig.~\ref{fig:fig1}(e)]. Previous works leveraged a wide range of photonic platforms, including metamaterials~\cite{benzaouia2024theory,baranov2017modifying,cortes2012quantum,molesky2020hierarchical,monticone2025nonlocality,de2025roadmap,yang2025orchestrating,lee2021angular}, waveguides~\cite{sheremet2023waveguide,sollner2015deterministic,ding2025purcell,gusken2023emission,hadad2020possibility,martin2025purcell,pak2022long}, cavities~\cite{solomon2024anomalous,vandrunen2024gain,saggio2024cavity,sakib2024purcell,pan2025room,ojambati2024few,chen2024scalable,gritsch2023purcell,sipahigil2016integrated}, plasmons~\cite{mueller2013asymmetric,rivera2016shrinking,lu2017dynamically,pick2017general,cai2024charge,cai2018photoluminescence,silveirinha2018fluctuation}, and photonic crystals~\cite{rugar2021quantum,yablonovitch1987inhibited,lyubarov2022amplified,park2025spontaneous}, to engineer spontaneous emission. By construction, these Purcell engineering platforms enhance/suppress photonic LDOS around atom resonance frequencies (optical or near-infrared for many emitters) through cavity resonances, photonic bands, and guided modes~\cite{barnes2020classical,sheremet2023waveguide,noda2007spontaneous}. In stark contrast, pure dephasing processes are generally connected to the broadband low-frequency (e.g., $\leq$MHz) fluctuations in the environments far off-resonant with the atom/spin transitions [see Fig.~\ref{fig:fig1}(e)]. The significant frequency and bandwidth mismatch necessitates the development of new structured photonic media with giant, broadband control over low-frequency components of photonic environments to tailor pure quantum dephasing dynamics.

In this paper, we introduce a nanophotonic approach to engineer pure quantum dephasing dynamics in quantum impurity systems. Unlike Purcell engineering platforms, our dephasing engineering leverages ultra-subwavelength spin noise metasurfaces which enable broadband control of low-frequency near-field EM environments qualitatively different from free space or macroscopic cavities. Our approach controls pure dephasing via metasurface-defined momentum filter functions for low-frequency evanescent waves in photonic environments, different from dephasing engineering based on dynamical decoupling (DD) pulses~\cite{cywinski2008enhance,yang2016quantum}. We experimentally demonstrate our nanophotonic quantum dephasing engineering using lithographically defined CoFeB metasurfaces on top of shallow nitrogen-vacancy (NV) center ensembles in diamond. Rather than modified spontaneous emission rates in Purcell engineering, we observe altered NV pure dephasing dynamics near different spin noise metasurfaces by measuring NV decoherence dynamics under Hahn echo. We further isolate metasurface-controlled dephasing from other intrinsic dephasing mechanisms (e.g., spin bath dephasing) by measuring the dephasing noise spectrum of NV ensembles with the spectral decomposition techniques based on Carr-Purcell-Meiboom-Gill (CPMG) pulse sequences~\cite{romach2015spectroscopy,bar2012suppression,ziffer2024quantum,liu2025quantum,alvarez2011measuring}. Our work establishes a new frontier in tailoring quantum light-matter interactions via nanophotonic structures, with potential applications in dephasing-assisted quantum energy transfer~\cite{rebentrost2009environment} for quantum chemistry, biology, and battery~\cite{shastri2025dephasing,lambert2013quantum,maier2019environment}. 

\section{Pure Quantum dephasing in photonic environments}
We start by describing the fundamentals of engineering quantum pure dephasing of a two-level system (TLS) using structured photonic environments. As sketched in Fig.~\ref{fig:fig1}(b), pure dephasing originates from random perturbations of the TLS resonance frequency by EM fluctuations in photonic environments~\cite{machado2023quantum,sun2025nanophotonic}. 
In pure dephasing, a TLS prepared in the coherent superposition state ($\frac{1}{\sqrt{2}}(|0\rangle + |1\rangle)$) gradually loses its phase coherence without photon emission/energy dissipation (Fig.~\ref{fig:fig1}(d)). This is in stark contrast to spontaneous emission processes, where a TLS prepared in the excited state emits a photon and loses its energy (see Fig.~\ref{fig:fig1}(a, c)). 

In the following, we consider a single magnetic TLS (spin qubit) coupled to arbitrary photonic environments characterized by the magnetic dyadic Green's function $\overleftrightarrow{G}_m$. We describe the pure dephasing dynamics through the dephasing function $\Phi(t)$ that governs the decay of the coherent superposition $\rho_{01}(t) = \rho_{01}(0) e^{-\Phi(t)}$, where $\rho_{01}(t)$ is the off-diagonal element of the single-qubit density matrix. The dephasing function $\Phi(t)$ due to magnetic field fluctuations in photonic environments is~\cite{sun2025nanophotonic},
\begin{align}
    &\begin{aligned}\label{dephsingf}
          &\Phi(\mathbf{r},t) = \frac{1}{\pi}\int_0^{\omega_c} d\omega \, F(\omega, t) J_{em}(\mathbf{r},\omega),
     \end{aligned}\\
    &\begin{aligned} 
          J_{em}(\mathbf{r},\omega)&=\frac{2\mu_0 \omega^2}{\hbar c^2} \coth{\frac{\hbar \omega}{2 k_B T}} \Big[\mathbf{m} \cdot 
         \mathrm{Im}  [\overleftrightarrow{G}_m(\mathbf{r},\omega)  +\overleftrightarrow{G}^\intercal_m(\mathbf{r},\omega)] \cdot \mathbf{m}^\dagger \Big]
         \label{Jem},
    \end{aligned}
\end{align} 
where $T$ is the environment temperature, $\mathbf{r}$ is the position of the TLS, $\mathbf{m}$ is the spin magnetic moment of the TLS, and $\omega_c$ is a cut-off frequency to avoid the ultraviolet divergence similar to Lamb shift calculations~\cite{sun2025nanophotonic}. As shown in Eq.~(\ref{dephsingf}), $\Phi(\mathbf{r},t)$ can be decomposed into the dephasing noise spectrum $J_{em}(\mathbf{r},\omega)$ and a frequency filter function $F(\omega, t)$. The dephasing noise spectrum $J_{em}(\mathbf{r},\omega)$ is associated with magnetic thermal and vacuum fluctuations in photonic environments. The frequency filter function $F(\omega, t)$ is determined by the microwave pulse sequence applied in the dephasing process, and is typically centered around $\mathrm{MHz}$ frequencies~\cite{cywinski2008enhance}. Therefore, from Eq.~(\ref{dephsingf}), pure dephasing processes are determined by the broadband low-frequency ($\sim\mathrm{MHz}$) EM fluctuations off-resonant with the TLS. 

For comparison, the well-known spontaneous emission rate of the magnetic TLS is~\cite{baranov2017modifying,sun2023limits},
\begin{equation}
    \Gamma(\mathbf{r}) = \frac{\mu_0\omega_{0}^2}{2\hbar c^2} \,  \big(\coth{\frac{\hbar \omega_0}{2 k_B T}} +1\big) \, \mathbf{m}^{eg}  \cdot \mathrm{Im} \, [\overleftrightarrow{G}_m  (\mathbf{r},\omega_0) +\overleftrightarrow{G}^\intercal_m(\mathbf{r},\omega_0)] \cdot \mathbf{m}^{eg\dagger},
\end{equation}
where $\omega_0$ is the resonance frequency and $\mathbf{m}^{eg}$ is the transition magnetic dipole moment of the TLS. In stark contrast to pure dephasing, it is clear that $\Gamma(\mathbf{r})$ is only determined by EM fluctuations at the TLS resonance frequency $\omega_0$. 

As shown in Fig.~\ref{fig:fig1}, prior works largely focused on modifying spontaneous emission rates $\Gamma$ by controlling the photonic LDOS at the emitter resonance frequency $\omega_0$~\cite{cortes2012quantum,baranov2017modifying}. In stark contrast, in this paper, our focus is engineering the dephasing function $\Phi(t)$ by tailoring the broadband low-frequency ($\sim$MHz) EM fluctuations in structured photonic environments.

\section{Spin noise metasurfaces for dephasing engineering} 

\begin{figure}[!t]
    \centering
    \includegraphics[width=3.6 in]{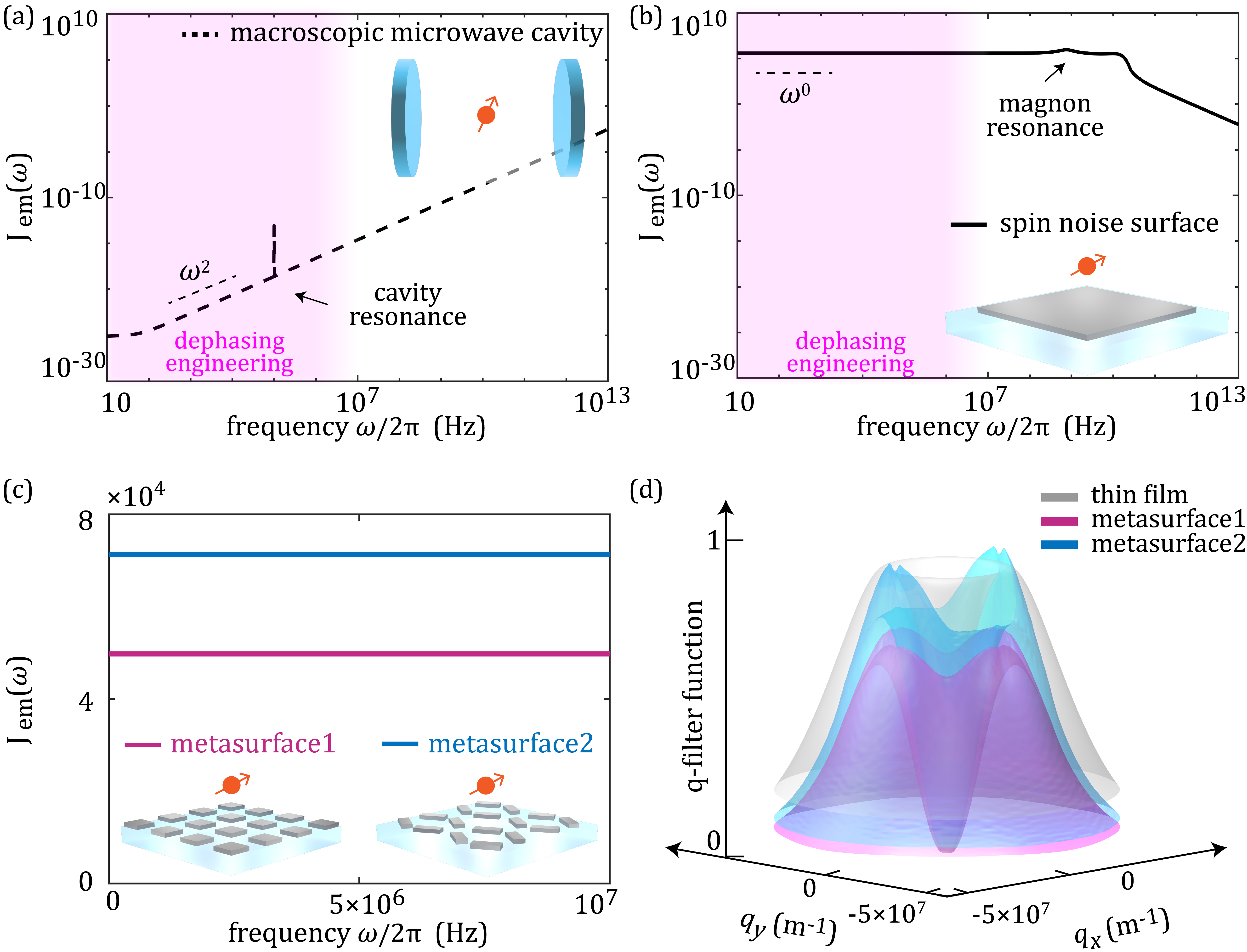}
    \caption{Spin noise metasurfaces for pure quantum dephasing engineering. (a, b) Low-frequency dephasing noise spectrum $J_{em}(\omega)$ in (a) macroscopic microwave cavities and (b) the near-field of spin noise surfaces. The broadband enhancement of $J_{em}(\omega)\sim\omega^0$ near spin noise surfaces provides photonic platforms for quantum pure dephasing engineering. (c) Engineered nanophotonic dephasing noise spectrum $J_{em}$ near two spin noise metasurfaces with the same filling factor and different geometries. (d) Spin noise metasurfaces control pure quantum dephasing via geometry-defined momentum space filters of low-frequency evanescent waves.} 
    \label{fig:fig2}
\end{figure}

To elucidate our approach for pure quantum dephasing engineering, we first unravel the properties of low-frequency magnetic field fluctuations in different photonic environments. Unlike the Purcell effect engineering feasible in various photonic platforms, we find that dephasing engineering requires ultra-subwavelength near-field nanophotonic environments. 

In Fig.~\ref{fig:fig2}(a,~b), we compare the low-frequency dephasing noise spectrum $J_{em}(\omega)$ in two classes of photonic structures, including the near-field of ferromagnetic materials and macroscopic microwave cavities. 
In Fig.~\ref{fig:fig2}(a), we plot the low-frequency $J_{em}(\omega)$ at the center of a single-mode microwave resonant cavity. We find that resonant cavities generally only modify $J_{em}(\omega)$ at cavity resonance frequencies $\omega_{cav}$, while $J_{em}(\omega \neq \omega_{cav})\sim\omega^2$ exhibits behaviors similar to free space at low frequencies. Therefore, although microwave resonant cavities are commonly used for Purcell engineering, they cannot provide effective broadband control over low-frequency $J_{em}(\omega)$ necessary for dephasing engineering. 

In Fig.~\ref{fig:fig2}(b), we demonstrate the $J_{em}(\omega)$ at distance $d=45\,\mathrm{nm}$ from a ferromagnetic \ce{CoFeB} thin film with $\sim 5$nm thickness. Here, we find that the near-field of ferromagnetic surfaces creates an ultra-subwavelength EM environment where the low-frequency EM noise is enhanced significantly in a broadband range and follows qualitatively different scaling $J_{em}(\omega) \sim \omega^0$ compared to free space or resonant cavities. This broadband enhancement of $J_{em}(\omega)$ originates from evanescent waves in the near-field with in-plane momenta $q \gg k_0$~\cite{ford1984electromagnetic}, and creates a unique photonic platform for controlling pure quantum dephasing processes.

We introduce ultra-subwavelength spin noise metasurfaces to effectively control the nanophotonic dephasing noise spectrum $J_{em}(\omega)$ in this ultra-subwavelength near-field regime. Previous works largely engineered near-field photonic environments at frequencies comparable to resonant modes of metasurfaces, e.g., surface lattice resonances~\cite{vecchi2009shaping,kravets2018plasmonic}, for spontaneous emission control. Here, we explore a unique regime of metasurface engineering, which focuses on low-frequency near-field
magnetic fluctuations ($\sim$MHz) far off-resonant with metasurface resonance. We show that spin noise metasurfaces introduce geometry-defined momentum filters for highly evanescent waves to control $J_{em}$. 

To make our approach explicit, we consider periodic metasurfaces made of thin ($\sim5$nm) ferromagnetic CoFeB with magnetic susceptibility $\overleftrightarrow{\chi}(\omega)$ following the Landau-Lifshitz-Gilbert (LLG) formula~\cite{cullity2011introduction} [Fig.~\ref{fig:fig2}(c)]. Since standard numerical methods (e.g., FEM, FDTD) become numerically ill-conditioned in the ultra-subwavelength near-field regime~\cite{zhu2014solution,sun2024computational}, we develop a self-consistent volume integral equation to capture the collective response of metasurfaces (see Methods and Supplemental S2). Physically, the Langevin spin noise $\langle \hat{S}_{f}^2\rangle \propto \mathcal{I}\mathit{m}\overleftrightarrow{\chi}(\omega) = (\overleftrightarrow{\chi}-\overleftrightarrow{\chi}^\dagger)/2i$ at metasurfaces generate near-field evanescent dipolar fields. Unlike thin films, highly evanescent fields with different momenta do not propagate independently in patterned metasurfaces, but can experience metasurface-defined couplings, which structure their contributions to low-frequency near-field magnetic fluctuations $\langle \hat{B}_{f}^2 \rangle \propto (\overleftrightarrow{G}_m - \overleftrightarrow{G}^\dagger_m)/2i$~\cite{buhmann2013dispersion}. We find the spatially averaged nanophotonic dephasing noise spectrum at distance $d$ near arbitrary spin noise metasurfaces (see Methods and Supplemental S2 for derivations),
\begin{align}\label{Gm_ms}
\begin{aligned}
    & J_{em}(d,\omega) = \frac{\mu_0}{\hbar \pi^2} \coth{\frac{\hbar \omega}{2 k_B T}} \,\mathbf{m} \cdot \mathrm{Re}\Big[ \int d^2\mathbf{q} \, \sum_{\mathbf{G},\mathbf{G'}}  \\ & \qquad \qquad \qquad \qquad \qquad \overleftrightarrow{G}_{0}(\mathbf{q},d) \cdot \mathcal{T}_{\mathbf{G}} \cdot \mathcal{I}\mathit{m} \big[ \overleftrightarrow{\chi}_{\mathbf{G},\mathbf{G'}} (\omega) \big] \cdot \mathcal{T}^\dagger_{\mathbf{G'}} \cdot \overleftrightarrow{G}^\dagger_{0}(\mathbf{q},d) \Big] \cdot \mathbf{m}^\dagger,
\end{aligned}
\end{align}
where $k_0=\omega/c$, $\mathbf{q}$ is the in-plane momentum of evanescent waves, and  $\mathbf{G}, \mathbf{G'}$ are the reciprocal lattice vectors of metasurfaces. Equation~(\ref{Gm_ms}) separates three physical ingredients, $\mathcal{I}\mathit{m} \overleftrightarrow{\chi}_{\mathbf{G},\mathbf{G'}}$ is the Fourier component of the periodic susceptibility set by the metasurface geometry, $\mathcal{T}_{\mathbf{G}}$ captures the couplings between evanescent waves with momenta $\mathbf{q}$ and $\mathbf{q+G}$ due to metasurfaces, and $\overleftrightarrow{G}_0(\mathbf{q})$ is the free space Green function that propagates structured spin noise at metasurfaces to $J_{em}$ in the near-field. 

To illustrate the metasurface effects, we show nanophotonic dephasing noise spectra $J_{em}(\omega)$ at $d=45\,\mathrm{nm}$ from two spin noise metasurfaces with similar filling factors in Fig.~\ref{fig:fig2}(c), where metasurface 1 consists of $150\,\mathrm{nm}$ square meta-atoms with $250\,\mathrm{nm}$ periodicity, and metasurface 2 contains four rotated $125\,\mathrm{nm}\times175\,\mathrm{nm}$ rectangular cells in one meta-atom with $500\,\mathrm{nm}$ periodicity. Both metasurfaces exhibit universal $J_{em} \sim \omega^0$, which can be derived from $\overleftrightarrow{\chi}(\omega) \sim \omega$ following LLG at low frequencies (see Supplemental S2), and $J_{em}(\omega)\sim \mathcal{I}\mathit{m} \overleftrightarrow{\chi}(\omega)/\omega \sim \omega^0$ based on Eq.~(\ref{Gm_ms}). We exploit this universal scaling to isolate metasurface-controlled nanophotonic dephasing noise in our experiments in the next sections. 

To reveal metasurface geometry effects, we demonstrate the momentum filters introduced by spin noise metasurfaces in Fig.~\ref{fig:fig2}(d). Here, for thin films and metasurfaces, the momentum filters peak at $q \gg k_0 \sim 0.01\mathrm
{m}^{-1}$, confirming that $J_{em}$ is dominated by contributions from highly evanescent waves. Meanwhile, unlike previous results near thin films~\cite{ford1984electromagnetic}, patterned spin noise metasurfaces couple evanescent waves with different momenta, and reshape their contributions to $J_{em}(\omega)$ through anisotropic momentum filters, as shown in Fig.~\ref{fig:fig2}(d). The resulting filters inherit the symmetry of metasurfaces, e.g., $C_4$ for metasurface 1 and $C_2$ for metasurface 2. Therefore, our nanophotonic engineering of pure quantum dephasing leverages metasurface-defined momentum filter functions, different from previous works using dynamical decoupling pulse engineering~\cite{cywinski2008enhance}. 

\section{Experimental Observation of engineered Spin dephasing dynamics} 

\begin{figure*}[!t]
    \centering
    \includegraphics[width = 4.6in]{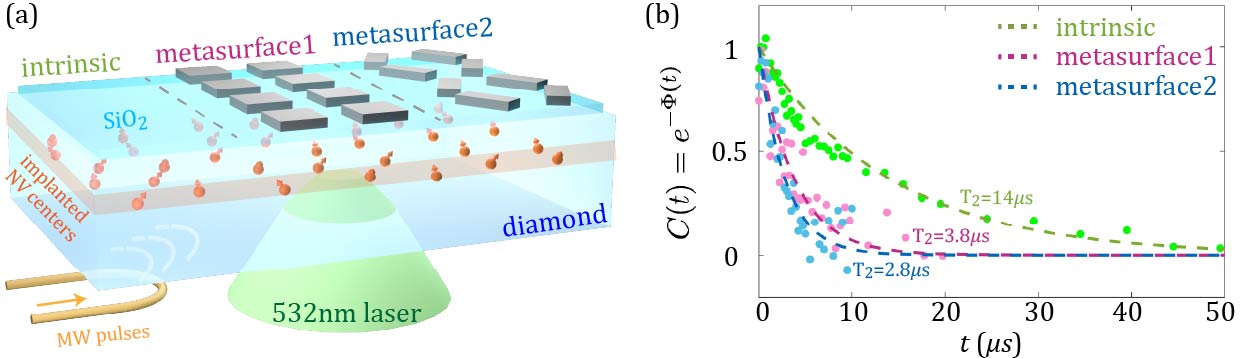}
    \caption{Observing engineered spin dephasing dynamics near ultra-subwavelength spin noise metasurfaces. (a) A schematic of a diamond device with two lithographically defined CoFeB spin noise metasurfaces and ion-implanted shallow NV center ensembles. (b) Measured NV decoherence dynamics under Hahn echo. The NV coherence function $C(t)=e^{-\Phi(t)}$ characterizes the dephasing dynamics near different spin noise metasurfaces. Decoherence time $T_2$ is defined by the dephasing function $\Phi(T_2)=1$.}
    \label{fig:fig3}
\end{figure*}

We experimentally demonstrate our nanophotonic quantum dephasing engineering using shallow nitrogen-vacancy (NV) center ensembles in diamond. Negatively charged NV centers have spin-triplet electronic ground states with spin sublevels $|m_s=0\rangle$ and $|m_s=\pm 1 \rangle$ separated by $2.87\,\mathrm{GHz}$. The coherent superposition of $|0\rangle$ and $|\pm 1 \rangle$ states has long spin coherence time at room temperatures with relatively well understood intrinsic decoherence mechanisms~\cite{sangtawesin2019origins,degen2017quantum,myers2014probing,bauch2020decoherence,de2010universal,hanson2006room,childress2006coherent,rondin2014magnetometry,du2017control}, which facilitate the isolation of photonic environment effects on NV spin dephasing. Unlike previous Purcell engineering experiments focusing on modified spontaneous emission rates $\Gamma$ in photonic environments~\cite{cortes2012quantum,baranov2017modifying,monticone2025nonlocality,de2025roadmap}, our dephasing engineering experiments focus on measuring the dephasing function $\Phi(t)$ of NV superposition states near ultra-subwavelength spin noise metasurfaces. 

To elucidate the influence of spin noise metasurfaces on NV spin dephasing, we fabricated a diamond device with three regions corresponding to different photonic environments, as shown in Fig.~\ref{fig:fig3}(a). In experiments, we created shallow NV ensembles with estimated depth around $\sim$$30\,\mathrm{nm}$ through ion-implantation. We designed and fabricated two ferromagnetic \ce{CoFeB} metasurfaces with similar filling factors using e-beam lithography and magnetron
sputtering on top of a deposited $\sim$15-$\mathrm{nm}$-thick \ce{SiO2} spacer layer on the diamond surface. The metasurface 1 [Fig.~\ref{fig:fig3}(a)] contains $150\,\mathrm{nm}\times150\,\mathrm{nm}$ \ce{CoFeB} meta-atoms with periodicity $250\,\mathrm{nm}\times250\,\mathrm{nm}$. The metasurface 2 [Fig.~\ref{fig:fig3}(a)] has periodicity $500\,\mathrm{nm}\times500\,\mathrm{nm}$ with each unit cell containing four $123\,\mathrm{nm}\times176\,\mathrm{nm}$ \ce{CoFeB} rectangular cells. The thickness of the deposited \ce{CoFeB} metasurfaces is around $5\,\mathrm{nm}$. 

We measure the spin coherence function $C(t)\approx e^{-\Phi(t)}$ of NV centers in different regions in our device by conducting standard spin Hahn-echo $T_2$ measurements using a home-built confocal microscopy setup~\cite{mahmud2025quantum}. In our experiment, we initialize the NV electron spin to the $|m_s=0\rangle$ state using a $532\,\mathrm{nm}$ green laser. We then apply a $\pi/2$ pulse to create coherent superposition $(|m_s=0\rangle + |m_s=1\rangle)/\sqrt{2}$ between different NV spin states. The superposition is then left to dephase for time $t/2$, followed by a $\pi$ pulse, and left to dephase for another time $t/2$. After the total dephasing time $t$, we apply another $\pi/2$ pulse to map the NV coherence function $C(t)$ to population differences for optical readout. By repeating the measurement and sweep $t$, we obtain the time evolution of the spin coherence function $C(t)$. Since our NV spin relaxation ($\sim102\,\mathrm{\mu s}$ near metasurface 1, $\sim70\,\mathrm{\mu s}$ near metasurface 2) is much slower than spin dephasing, we neglect the relaxation contribution to decoherence and find $C(t)\approx e^{-\Phi(t)}$.

In Fig.~\ref{fig:fig3}(b), we demonstrate the measured NV dephasing processes $e^{-\Phi(t)}$ in different regions in our device without external magnetic fields. We define the decoherence time $T_2$ as $\Phi(T_2)=1$. In our device, the NV electron spin dephasing is dominated by two noise sources, including the nanophotonic dephasing due to magnetic fluctuations in near-field EM environments and spin bath dephasing due to flip-flop processes of di-vacancy \ce{V2} centers created in the vicinity of NVs during
the ion-implantation process~\cite{favaro2017tailoring}. 
In the region far from spin noise metasurfaces, the NV spin dephasing in our device is dominated by intrinsic noise associated with the di-vacancy center \ce{V2} spin bath. We find that the NV spin decoherence time $T_2$ induced by this intrinsic noise is $\sim 14\,\mathrm{\mu s}$. Meanwhile, we observe significant changes in the NV spin decoherence time $T_2$ near spin noise metasurfaces. Here, spin noise metasurfaces introduce additional nanophotonic dephasing dependent on the metasurface structures. We find $T_2\approx 3.8\,\mathrm{\mu s}$ near the metasurface 1 and $T_2\approx 2.8\,\mathrm{\mu s}$ near the metasurface 2. Our measurements demonstrate the nanophotonic engineering of the quantum pure dephasing dynamics using spin noise metasurfaces. 

\section{Isolating Metasurface-Controlled Nanophotonic Dephasing Noise Spectrum}

\begin{figure}[!t]
    \centering
    \includegraphics[width=5.4 in]{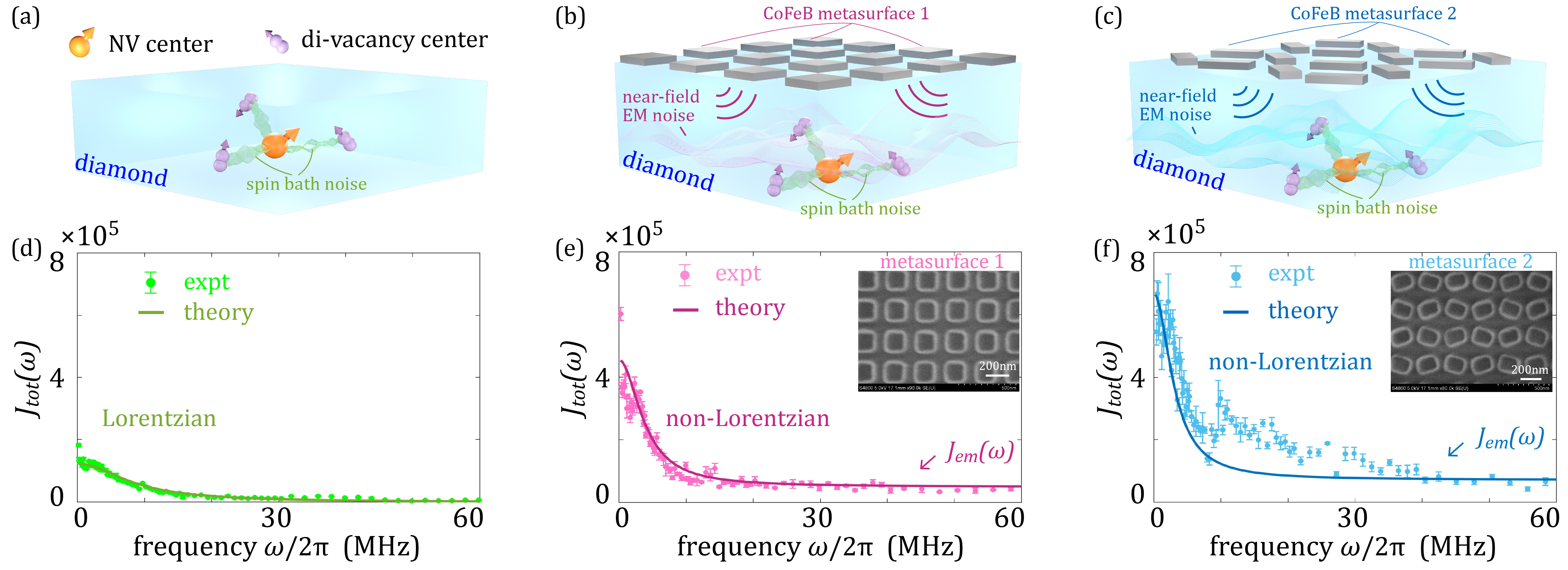}
    \caption{Isolating metasurface-engineered dephasing noise spectrum $J_{em}(\omega)$ from other intrinsic dephasing noise. (a-c) Schematics of different dephasing noise sources in the diamond NV device, including nanophotonic dephasing $J_{em}(\omega)$ associated with spin noise metasurfaces and intrinsic dephasing $J_{s-b}(\omega)$ connected to the \ce{V2} spin bath. (d-f) Total NV dephasing noise spectrum $J_{tot}(\omega)=J_{em}(\omega)+J_{s-b}(\omega)$ near different metasurfaces measured with the spectral decomposition techniques based on CPMG pulse sequences. Metasurface-engineered dephasing noise $J_{em}(\omega)$ leads to the non-Lorentzian behaviors and slowly-decaying tails at higher frequencies ($\geq10$MHz) in (e) and (f). The solid lines are the theory fitting to the experimental data using Eqs.~(\ref{Jsb},~\ref{J_ms}). Insets show the SEM images of CoFeB metasurfaces.}
    \label{fig:fig4}
\end{figure}

To isolate nanophotonic dephasing from other intrinsic dephasing mechanisms (e.g., \ce{V2} spin bath dephasing), we now study the dephasing noise spectrum $J_{tot}(\omega)$ for NV centers in different regions in our device [Fig.~\ref{fig:fig4}(a-c)]. As shown in Fig.~\ref{fig:fig2}(b), the nanophotonic dephasing noise spectrum $J_{em}(\omega)$ induced by magnetic fluctuations in nanophotonic environments shows negligible dependence on frequencies $J_{em}(\omega)\sim\omega^0$. In contrast, spin bath dephasing noise spectrum $J_{s-b}(\omega)$ originates from interactions between NV centers and vicinity di-vacancy center \ce{V2} bath~\cite{favaro2017tailoring}, and follows Lorentzian behaviors with fast decay at MHz frequencies~\cite{favaro2017tailoring}. We leverage the qualitatively different behaviors of nanophotonic dephasing noise spectrum $J_{em}$ and spin bath dephasing noise spectrum $J_{s-b}$ to isolate dephasing noise originating from different noise sources.

We measure the dephasing noise spectrum $J_{tot}(\omega)$ using the spectral decomposition technique based on the CPMG pulse sequence~\cite{romach2015spectroscopy,bar2012suppression}. The N pulse CPMG sequence contains N $\pi$-pulses and creates narrow band frequency filter functions $F_N(\omega,t)$ with center frequencies $\omega_N$ shifting with $N$~\cite{cywinski2008enhance} (see Methods). From Eq.~(\ref{dephsingf}), we can then extract the noise spectrum $J_{tot}(\omega_N)$ at the center frequencies $\omega_N$ of the filter functions through measuring the spin coherence function $C_N(t) \approx e^{-\Phi_N(t)}$ under N pulse CPMG sequences, i.e., $J_{tot}(\omega_N)=-\pi\ln{C_N(t)}/t$. The spot size in our confocal microscopy setup is around $\sim$$500\,\mathrm{nm}$ larger than the periodicity of the spin noise metasurfaces. Therefore, our measured $J_{tot}(\omega)$ represents the spatially and ensemble averaged noise spectrum near the spin noise metasurfaces.

In Fig.~\ref{fig:fig4}(d-f), we demonstrate the measured dephasing noise spectrum $J_{tot}(\omega)$ for NVs in different regions in our device. Here, we find that $J_{tot}(\omega)$ for NVs away from the spin noise metasurfaces follows the Lorentzian behavior expected for intrinsic noise. In contrast, near spin noise metasurfaces, we observe that the dephasing noise spectrum $J_{tot}(\omega)$ exhibits non-Lorentzian behaviors qualitatively different from the intrinsic spin bath noise. Notably, $J_{tot}(\omega)$ near spin noise metasurfaces exhibits slowly-decaying tails at higher frequencies sensitive to the geometries of spin noise metasurfaces. 

We interpret the above behaviors of $J_{tot}(\omega)$ using a two-bath model, where the NV ensembles are coupled to both the intrinsic spin bath $J_{s-b}(\omega)$ and the near-field photon bath $J_{em}(\omega)$ controlled by spin noise metasurfaces. Away from the metasurfaces, the NV dephasing noise spectrum (Fig.~\ref{fig:fig4}(d)) is dominated by contributions from the di-vacancy centers \ce{V2} spin bath, 
\begin{equation}\label{Jsb}
    J_{s-b}(\omega)=\frac{\Delta^2 \tau_c}{\pi}\frac{1}{1+(\omega \tau_c)^2},
\end{equation}
where $\Delta$ is the coupling strength and $\tau_c$ is the bath correlations of the di-vacancy center spin bath. Fitting with the experimental data in Fig.~\ref{fig:fig4}(d) gives $\Delta \approx 4.5\,\mathrm{MHz}, \tau_c \approx 19\,\mathrm{ns}$ comparable to previously reported values for \ce{V2} spin bath~\cite{favaro2017tailoring}. This supports the interpretation that the intrinsic noise in our device primarily comes from the \ce{V2} spin bath relevant to implantation-induced defects~\cite{favaro2017tailoring}. 

Meanwhile, near spin noise metasurfaces, metasurface-controlled nanophotonic dephasing noise $J_{em}(\omega)$ contributes to the total dephasing noise spectrum, 
\begin{equation}\label{J_ms}
    J_{tot}(\omega)=J_{s-b}(\omega)+J_{em}(\omega)=\frac{\Delta'^2 \tau_c'}{\pi}\frac{1}{1+(\omega \tau_c')^2}+J_{em}(\omega).
\end{equation}
where $J_{em}(\omega)$ denotes the spatially and ensemble averaged nanophotonic dephasing noise spectrum for the NV ensemble in the readout laser spot, $\Delta'$ and $\tau_c'$ are the coupling strength and bath correlations of the \ce{V2} spin bath near the spin noise metasurfaces. From Eq.~\eqref{J_ms}, we isolate the nanophotonic dephasing and spin bath dephasing in the noise spectrum in Fig.~\ref{fig:fig4}(e,f) through their distinct frequency dependences. We find the Lorentzian spin-bath term $J_{s-b}$ dominates at low frequencies and rapidly decays at higher frequencies, whereas the metasurface contribution is weakly frequency dependent in the MHz range, with $J_{em}\sim\omega^0$ from Fig.~\ref{fig:fig2}, and dominates at higher frequencies in Fig.~\ref{fig:fig4}(e,f). To interpret the measured spectra, we use $J_{em}(\omega)$ in Fig.~\ref{fig:fig2}(c) from the metasurface calculations, where we take CoFeB magnetic parameters comparable to previously reported values~\cite{tang2017thickness,thiruvengadam2022anisotropy} (see Methods) and fit the high-frequency tail of the measured spectra. In fitting Eq.~(\ref{J_ms}), $\Delta'$ is kept close to the intrinsic spin-bath coupling scale, as it is mostly set by the static dipole-dipole interactions between NV and \ce{V2} spins, while $\tau_c'$ is allowed to vary because it is related to local spin bath dissipation~\cite{bauch2020decoherence} modified by spin noise metasurfaces (see Supplement S3 for more details). Together, we find qualitative match between Eq.~(\ref{J_ms}) and our experimental measurements near both metasurfaces, where we take  $\Delta'\approx5.6\,\mathrm{MHz}, \tau_c' \approx 41\,\mathrm{ns}$ in Fig.~\ref{fig:fig4}(e), and $\Delta'\approx 6 \,\mathrm{MHz}, \tau_c' \approx 51\,\mathrm{ns}$ in Fig.~\ref{fig:fig4}(f). We note that our approach isolates contributions from the nanophotonic dephasing and spin bath dephasing, and explains our observed slowly-decaying $J_{tot}(\omega)$ at higher frequencies near spin noise metasurfaces in Fig.~\ref{fig:fig4}. 

\section{Discussion}
We have theoretically proposed and experimentally demonstrated a nanophotonic approach to engineer quantum pure dephasing, a paradigm complementary to prior Purcell engineering of spontaneous emission,  using lithographically defined ultra-subwavelength CoFeB metasurfaces and shallow NV centers. Our approach can be extended to other qubit platforms, and can have potential applications in programmable open quantum systems and dephasing-assisted quantum energy transfer~\cite{rebentrost2009environment,plenio2008dephasing}. Despite the negative influence of dephasing on quantum information, previous works have identified that pure dephasing can be crucial for enhancing energy transfer efficiencies and rates in disordered quantum systems, leading to dephasing-assisted quantum transport regimes important for quantum chemistry~\cite{blach2025environment}, quantum biology~\cite{lambert2013quantum}, quantum thermodynamics~\cite{mendoza2013heat}, and quantum batteries~\cite{shastri2025dephasing}. Our work thus provides a new non-invasive, lithographically controlled approach different from microwave pulse engineering for dephasing control, paving the way for developing future open quantum systems tailored by structured electromagnetic environments. 

\section{Methods}
\subsection{Spectral Decomposition Techniques}
\begin{figure}[!h]
    \centering
    \includegraphics[width = 3.2in]{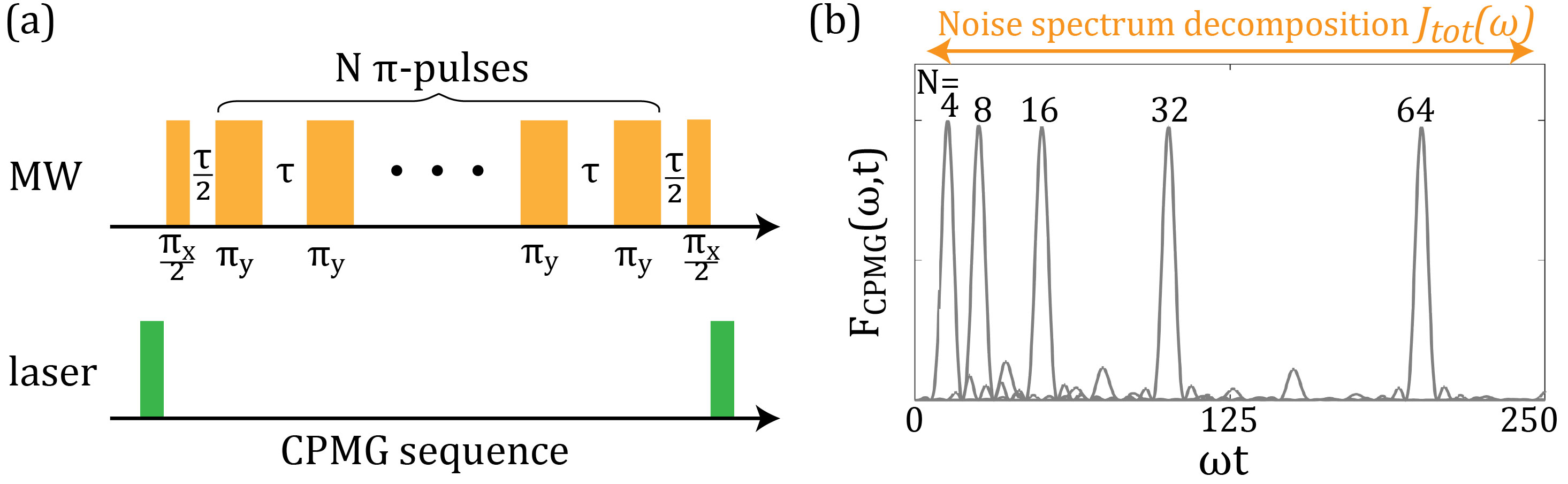}
    \caption{CPMG pulse sequences and spectral decomposition of $J_{tot}(\omega)$. (a) A schematic of microwave (MW) pulses and laser pulses for the CPMG pulse sequence measurements. (b) The $N$ pulse CPMG sequence creates a frequency filter function $F_{N}(\omega,t)$ centered at $\omega_N$ for spectral decomposition.}
    \label{fig:fig5}
\end{figure}

We measured the NV dephasing noise spectrum $J_{tot}(\omega)$ in Fig.~\ref{fig:fig4} using dynamical decoupling spectral decomposition~\cite{bar2012suppression,myers2014probing} based on the Carr-Purcell-Meiboom-Gill (CPMG) pulse sequences, as shown in Fig.~\ref{fig:fig5}(a). An $N$ pulse CPMG sequence contains $N$ equally spaced $\pi$ pulses during a total evolution time $t$. Physically, these $\pi$ pulses periodically invert the NV phase accumulation, averaging out noise at certain frequencies. This creates frequency filter functions $F_{N}(\omega,t)$, which are $\delta$-like functions centered around $\omega_N\sim\pi N/t$, as shown in Fig.~\ref{fig:fig5}(b)~\cite{cywinski2008enhance}. Therefore, under an $N$ pulse CPMG sequence, we have the NV dephasing function $\Phi_N(t) \approx t J_{tot}(\omega_N)/\pi$. In our experiments, we measure the NV spin coherence function $C(t)=e^{-\Phi_N(t)}$ under systematically varying $N$, and reconstruct $J_{tot}(\omega)\propto -\pi \ln{C(t)}/t$ over the MHz frequency range relevant to NV pure quantum dephasing.

\subsection{Numerical Simulations of Spin Noise Metasurfaces}
For photonic engineering of pure quantum dephasing dynamics, the relevant EM wavelength at MHz frequencies ($\sim$m) exceeds the metasurface feature sizes and NV-metasurface distance ($\sim 100\,\mathrm{nm}$) by several orders of magnitude, positioning the problem in the ultra-subwavelength near-field regime where common computational methods in quantum photonics (e.g., finite element (FEM) and finite difference time domain (FDTD)) become numerically ill-conditioned~\cite{zhu2014solution,sun2024computational}. Here, we introduce a volume integral equation (VIE) framework with fluctuational magnetization sources in reciprocal space, as a natural method to capture ultra-subwavelength spin noise metasurface physics and model the nanophotonic dephasing noise spectrum $J_{em}(\omega)$.

The structured fluctuating magnetization $\mathbf M(\mathbf r,\omega)$ in the metasurfaces follows the self-consistent equation,
\begin{equation}\label{VIE_r}
\mathbf M(\mathbf r,\omega)
= 
\mathbf M^{\rm fl}(\mathbf r,\omega) -
\overleftrightarrow{\chi}(\mathbf r,\omega) \cdot 
\overleftrightarrow{N}\cdot 
\mathbf M(\mathbf r,\omega) 
 + 
\overleftrightarrow{\chi}(\mathbf r,\omega)\cdot 
\int_{V_{ms}} d^3\mathbf r'\,
\overleftrightarrow G_0(\mathbf r,\mathbf r',\omega)\cdot 
\mathbf M(\mathbf r',\omega),
\end{equation}
where $\mathbf{M}^{\rm fl}$ is the Langevin fluctuating magnetization sources, $\overleftrightarrow G_0$ is the free-space magnetic dyadic Green function~\cite{novotny2012principles,khandekar2020new,liu2015near}, and $\overleftrightarrow{N}$ is the self-demagnetization tensor~\cite{biehs2021near}. Since our CoFeB metasurface thickness is much smaller than meta-atom sizes, we take $\overleftrightarrow{N}\approx \mathrm{diag([0,0,1])}$. The magnetic susceptibility $\overleftrightarrow{\chi}$ is periodically modulated by the metasurface geometry with the Fourier expansion, $\overleftrightarrow{\chi}(\boldsymbol{\rho},z,\omega)
=
\sum_{\mathbf G}
\overleftrightarrow{\chi}_{\mathbf G}(z,\omega)
e^{i\mathbf G\cdot\boldsymbol{\rho}}$, where $\boldsymbol{\rho}=(x,y)$, and $\mathbf G$ are reciprocal lattice vectors.

For simplicity, we suppress the explicit z-dependence in Eq.~(\ref{VIE_r}) appropriate for the $\sim$5nm thick CoFeB layers used in our experiments. We provide the full treatment of Eq.~(\ref{VIE_r}) in the Supplement S2. Within this approximation, Fourier transform of Eq.~(\ref{VIE_r}) with respect to the in-plane coordinates $\boldsymbol{\rho}$ gives,
\begin{equation}\label{eq:continuous_z_vie}
\mathbf M_{\mathbf G}(\mathbf q,\omega)
=
\mathbf M^{\rm fl}_{\mathbf G}(\mathbf q,\omega) +
\sum_{\mathbf G'}
\overleftrightarrow{\chi}_{\mathbf G'-\mathbf G}(\omega) \, \cdot 
 \Big[ 
\overleftrightarrow{G}_0(\mathbf q - \mathbf G',\omega)\cdot 
\mathbf M_{\mathbf G'}(\mathbf q,\omega)  - \overleftrightarrow{N}\cdot \mathbf M_{\mathbf G'}(\mathbf q,\omega) \Big],
\end{equation}
where $\mathbf M_{\mathbf G}(\mathbf q,\omega)
=
\mathbf M(\mathbf q - \mathbf G,\omega)$. From Eq.~\eqref{eq:continuous_z_vie}, we define a coupling kernel $\mathcal T_{\mathbf G\mathbf G'}$ that captures the metasurface-defined couplings of highly evanescent waves with different momenta and connects the dressed magnetization $\mathbf M_{\mathbf G}$ to $\mathbf M^{\mathrm{fl}}_{\mathbf G'}$ via  
$\mathbf M_{\mathbf G}(\mathbf q,\omega)
=
\sum_{\mathbf G'}
\mathcal{T}_{\mathbf G\mathbf G'}(\mathbf q,\omega)
\mathbf M^{\rm fl}_{\mathbf G'}(\mathbf q,\omega)$, and  $[\mathcal{T}^{-1}]_{\mathbf{G}\mathbf{G'}} = \delta_{\mathbf{G},\mathbf{G'}} I_3 - \chi_{\mathbf{G}'-\mathbf{G}}\cdot (\overleftrightarrow{G}_0(\mathbf{q}-\mathbf{G'})-\overleftrightarrow{N})$. In the ultra-subwavelength regime, obtaining $\mathcal{T}_{\mathbf{G}\mathbf{G'}}$ with direct inversion is computationally difficult. We therefore use an iterative solver to calculate $\mathcal{T}_{\mathbf{G}\mathbf{G'}}$.  

The final nanophotonic dephasing noise spectrum follows the reciprocal space VIE solutions. The FDT relates Langevin magnetization sources $\mathbf M^{\rm fl}$ to the dissipative susceptibility $\mathcal{I}\mathit{m}\,\overleftrightarrow{\chi}_{\mathbf G,\mathbf G'}(\omega) = 
\mathcal{I}\mathit{m}\,\overleftrightarrow{\chi}_{\mathbf G'-\mathbf G}(\omega)$. The metasurface couplings $\mathcal T_{\mathbf G}$ convert these fluctuating sources into the dressed magnetization noise $\mathbf M$, and the free space Green function $\overleftrightarrow G_0(\mathbf q,d,\omega)$ propagates this dressed noise to near-field magnetic fluctuations at NVs at distance $d$ from metasurfaces. Since our laser spot $(\sim500\,{\rm nm})$ is larger than the metasurface period, the experimental measurements correspond to spatially averaged nanophotonic dephasing noise spectrum $J_{\rm em}(d,\omega)$ in Eq.~\eqref{Gm_ms} in the main text with $\mathcal{T}_{\mathbf{G}}=\mathcal{T}_{\mathbf 0,\mathbf{G}}$.

In experiments, the measured signal comes from an NV ensemble with four crystallographic orientations. For Fig.~\ref{fig:fig2}, we further perform the ensemble average,
\begin{align}
\begin{aligned}
    J_{em}^{ens}(d,\omega)=\frac{1}{4} \sum_{i=1}^4 J_{em}(d,\mathbf{m}_i,\omega),
\end{aligned} 
\end{align} 
with $\mathbf{m}_i$ along four NV axes $(\cos{\alpha_i}\sin{\theta},\sin{\alpha_i}\sin{\theta},\cos{\theta}),\, i=1,\cdots,4$. For our diamond cut, we take $\theta = 54.7\degree$ and  $\alpha_i=\alpha_0+(i-1)\pi/2$. For Fig.~\ref{fig:fig2}(c), the CoFeB susceptibility is evaluated using the LLG formula with saturation magnetization $\mu_0 M_s=12500\,\mathrm{G}$, effective magnetization $\mu_0 M_{eff}=12500\,\mathrm{G}$, Gilbert damping factor $\alpha=0.019$, and in-plane uniaxial anisotropy field $H_k=5\,\mathrm{G}$, corresponding to previously reported CoFeB parameters~\cite{tang2017thickness,thiruvengadam2022anisotropy}. The in-plane easy-axis direction is determined by fitting to the experimental noise spectra in Fig.~\ref{fig:fig4}. For the momentum-filter plots in Fig.~\ref{fig:fig2}(d), we  use an in-plane isotropic magnetic response to isolate the geometry-defined momentum filters, excluding contributions from magnetic anisotropy in materials (see Supplement S2).

\subsection{Confocal Microscopy Setup and Sample Preparation}
The confocal microscopy used in this paper follows the setup used in our previous work~\cite{mahmud2025quantum}.

The device was fabricated on a $500\,\mathrm{\mu m}$ thick optical-grade CVD diamond substrate with $\langle 100 \rangle$ orientation and natural carbon isotope abundance. Shallow NV ensembles were formed by ${^{15}N}$ ion implantation at $20$keV with a dose of $10^{12}$cm$^{-2}$, giving an estimated NV depth of $\sim30\,\mathrm{nm}$. A $15\,\mathrm{nm}$ amorphous \ce{SiO2} spacer was deposited on the diamond surface to avoid direct electron transport and preserve NV properties in the diamond lattice. Periodic \ce{Co60Fe20B20} spin noise metasurfaces were then patterned by electron-beam lithography and magnetron sputtering, followed by a $12\,\mathrm{nm}$ alumina capping layer sputtered inside the same
chamber without breaking the vacuum. The CoFeB thickness was $\sim5\,\mathrm{nm}$, and the fabricated metasurface patterns were characterized by scanning electron microscopy. Full fabrication procedures are provided in Supplement S1.

\begin{backmatter}
\bmsection{Funding}
This work was supported by the Army Research Office under Grant No. W911NF-21-1-0287.

\bmsection{Disclosures}
The authors declare no conflicts of interest.

\bmsection{Data availability} Data underlying the results presented in this paper are not publicly available at this time but may be obtained from the authors upon reasonable request.

\bmsection{Supplemental document}
See supplemental document for supporting content.

\end{backmatter}

\bibliography{reference}

\end{document}


\preprint{APS/123-QED}

\title{Supplementary Material for `Beyond the Purcell Effect: Controlling Pure Quantum Dephasing with Spin Noise Metasurfaces'}

\author{Wenbo Sun}
\altaffiliation{equal contribution}
\affiliation{Elmore Family School of Electrical and Computer Engineering, Purdue University, West Lafayette, Indiana 47907, USA}
\affiliation{Birck Nanotechnology Center, Purdue University, West Lafayette, Indiana 47907, USA}
\author{Shoaib Mahmud}
\altaffiliation{equal contribution}
\affiliation{Elmore Family School of Electrical and Computer Engineering, Purdue University, West Lafayette, Indiana 47907, USA}
\affiliation{Birck Nanotechnology Center, Purdue University, West Lafayette, Indiana 47907, USA}
\author{Wei Zhang}
\altaffiliation{equal contribution}
\affiliation{Elmore Family School of Electrical and Computer Engineering, Purdue University, West Lafayette, Indiana 47907, USA}
\affiliation{Birck Nanotechnology Center, Purdue University, West Lafayette, Indiana 47907, USA}
\author{Runwei Zhou}
\affiliation{Elmore Family School of Electrical and Computer Engineering, Purdue University, West Lafayette, Indiana 47907, USA}
\author{Pronoy Das}
\affiliation{Elmore Family School of Electrical and Computer Engineering, Purdue University, West Lafayette, Indiana 47907, USA}
\affiliation{Birck Nanotechnology Center, Purdue University, West Lafayette, Indiana 47907, USA}
\author{Dan Jiao}
\affiliation{Elmore Family School of Electrical and Computer Engineering, Purdue University, West Lafayette, Indiana 47907, USA}
\author{Zubin Jacob}
\email{zjacob@purdue.edu}
\affiliation{Elmore Family School of Electrical and Computer Engineering, Purdue University, West Lafayette, Indiana 47907, USA}
\affiliation{Birck Nanotechnology Center, Purdue University, West Lafayette, Indiana 47907, USA}

\date{\today}

\maketitle

\section{Sample Fabrication}
\begin{figure*}
    \centering
    \includegraphics[width = 5in]{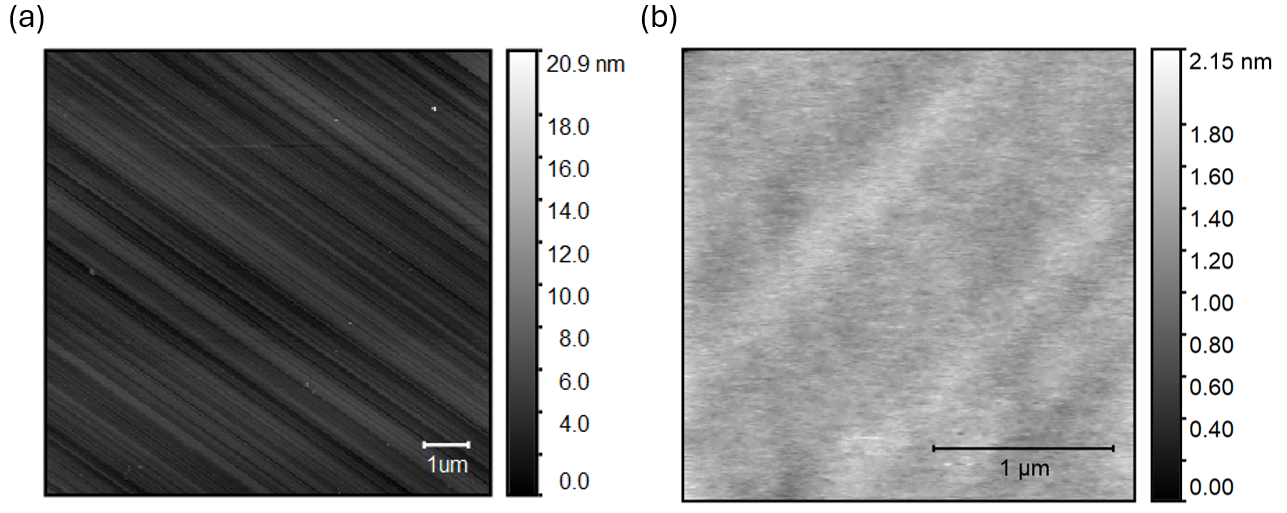}
    \caption{Diamond surface topography characterization using atomic force microscopy. (a) Surface roughness of the diamond substrate before etching was around 1nm rms. (b) After etching, the surface roughness was reduced to about 300pm rms.}
    \label{fig:figS1}
\end{figure*}

We fabricated nanoscale periodic spintronic meta-atoms directly on the surface of the NV-implanted diamond to investigate the near-field spin noise effects on the defect centers. The substrate we used is a 500 $\mu m$ thick optical-grade chemical vapor deposition-grown diamond slab (Element Six) with dimensions of 4.5 mm by 4.5 mm and a lattice orientation of $\langle 100 \rangle$. The diamond sample has a natural isotopic abundance of carbon of about $99\%$ ${^{12}C}$ and $1\%$ ${^{13}C}$.  First, we etched $5\mu m$ on each surface of the substrate in Plasma-Therm Apex SLR inductively coupled plasma etcher using \ce{Ar}, \ce{O2}, and \ce{Cl2}. This process released the surface tension of the diamond slab during growth and polishing. It also further reduced the surface roughness to below $500 pm$ rms for the subsequent fabrication as shown in Figure~\ref{fig:figS1}.  The sample was then soaked in piranha solution, followed by nitric acid to dissolve organic contamination and graphitize diamond layers. Inorganic residues on the sample surface were washed away in the standard solvent cleaning procedure. A layer of NVs $\sim$30 nm under the surface was created by the implantation of Nitrogen-15 ions with 20 keV energy and $10^{12}$/cm$^{2}$ density.

\begin{figure*}
    \centering
    \includegraphics[width = 6in]{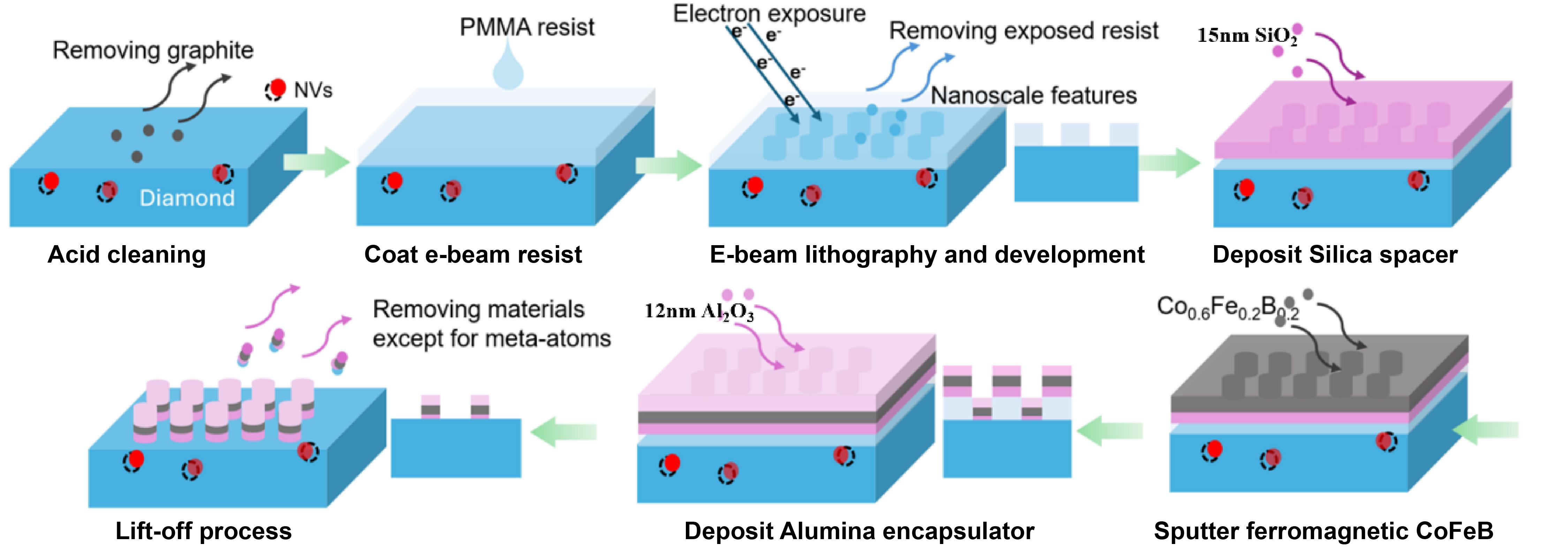}
    \caption{Fabrication procedures for our diamond device with spin noise metasurfaces.}
    \label{fig:figS2}
\end{figure*}

Before growing the magnetic material on diamond, we deposited 15nm amorphous silica spacer directly on the surface of diamond to avoid direct electron transport and preserve NVs' properties in the diamond lattice. We started by spinning 950PMMA A4 electron beam resist on the diamond substrate with 3500 rpm and a 180 $^\circ C$ afterbake on a hot plate. The designed metasurface patterns were then patterned on the substrate by the standard electron beam lithography exposure in a JEOL JBX-8100 FS E-beam writer using 2nA current and 100 keV beam energy. The magnetic material we used is \ce{Co60Fe20B20}, and a stoichiometric material source was installed into a PVD sputtering chamber with a base pressure of $10^{-6}$ Torr for magnetron sputtering. The CoFeB material was sputtered for a calibrated time of 9mins and 18s, immediately followed by a 12nm alumina capping layer sputtered inside the same chamber. After the growing process, the diamond sample was lifted off in the acetone bath and imaged using scanning electron microscopy in a Thermo Fisher Scientific Apreo SEM. The full fabrication procedures for our devices are demonstrated in Fig.~\ref{fig:figS2}.

\section{Numerical Simulations of Spin Noise Metasurfaces}
In this section, we present the derivations and numerical implementations of our fluctuational volume integral equation (VIE) method for the ultra-subwavelength spin noise metasurfaces and the near-field nanophotonic dephasing noise spectrum $J_{em}(\omega)$ in the main text. 

\subsection{Fluctuational Volume Integral Equation for Spin Noise Metasurfaces}

The photonic engineering of pure quantum dephasing studied here lies far outside the usual wavelength-scale nanophotonic simulations based on finite element (FEM) or finite-difference time-domain (FDTD) methods~\cite{zhu2014solution,sun2024computational}. At MHz frequencies, the electromagnetic wavelength ($>$m) is more than 7 orders of magnitude larger than both the metasurface feature size and the qubit-metasurface separation. In this ultra-subwavelength near-field regime, we develop a VIE with fluctuational sources in reciprocal space as a natural route to capture spin noise metasurface physics. We start from the real space VIE for the structured magnetization inside metasurfaces, 
\begin{align}\label{S2_real_space_VIE}
\mathbf{M}(\mathbf{r},\omega)
=
\mathbf{M}^{\rm fl}(\mathbf{r},\omega)
-
\overleftrightarrow{\chi}(\mathbf{r},\omega)
\cdot
\overleftrightarrow{N}
\cdot
\mathbf{M}(\mathbf{r},\omega)
+
\overleftrightarrow{\chi}(\mathbf{r},\omega)
\cdot
\int_{V_{ms}} d^3r'\,
\overleftrightarrow{G}_0(\mathbf{r},\mathbf{r}',\omega)
\cdot
\mathbf{M}(\mathbf{r}',\omega),
\end{align}
where $\mathbf{M}^{\rm fl}$ is the Langevin fluctuating magnetization source, $\overleftrightarrow{\chi}(\mathbf{r})$ is the magnetic susceptibility of the patterned CoFeB, $\overleftrightarrow{G}_0$ is the free-space magnetic dyadic Green function that connects magnetization at metasurfaces to near-field magnetic fields, and $\overleftrightarrow{N}$ is the local demagnetization tensor introduced by principal-value regularization of the singular self-interaction~\cite{biehs2021near}. For the ultrathin CoFeB layers used in our experiments, we take
\begin{align}
\overleftrightarrow{N}
\simeq
\begin{bmatrix}
    0 & 0 & 0\\
    0 & 0 & 0\\
    0 & 0 & 1
\end{bmatrix},
\end{align}
and the free-space dyadic Green function $\overleftrightarrow{G}_0$~\cite{novotny2012principles},
\begin{align}
\begin{aligned}
\overleftrightarrow{G}_0(\mathbf{r},\mathbf{r}',\omega)=\int \frac{d^2 \mathbf{q}}{4\pi^2} e^{i\mathbf{q}\cdot(\boldsymbol{\rho}-\boldsymbol{\rho}')}\overleftrightarrow{G}_0(\mathbf{q},z,z',\omega),
\qquad 
\overleftrightarrow{G}_0(\mathbf{q},z,z',\omega) =
\frac{i}{2} \frac{1}{k_z}
\begin{pmatrix}
k_0^2-q_x^2 & -q_xq_y & -s q_xk_z\\
-q_xq_y & k_0^2-q_y^2 & -s q_yk_z\\
-s q_xk_z & -s q_yk_z & k_0^2-k_z^2
\end{pmatrix} 
e^{i k_z |z-z'|},
\end{aligned}
\end{align}
where $\mathbf{q}$ is the in-plane momentum, $k_0=\omega/c$, $k_z=\sqrt{k_0^2-q^2}$, and $s={\rm sgn}(z-z')$. $\boldsymbol{\rho}=(x,y)$ represents the in-plane coordinates and $\mathbf{r}=(\boldsymbol{\rho},z)$.

In the ultra-subwavelength near-field regime, the dephasing noise is dominated by evanescent magnetic fields with in-plane momenta $q \gg k_0$. We therefore formulate Eq.~\eqref{S2_real_space_VIE} in a mixed $(\mathbf{q},z)$ representation by Fourier transforming only the in-plane coordinates $\boldsymbol{\rho}$ while keeping the $z$ dependence explicit. This representation naturally resolves the dominant evanescent wave contributions and directly reveals how the metasurface geometry structures these evanescent wave contributions via momentum filters for $J_{em}(\omega)$. In Eq.~\eqref{S2_real_space_VIE}, the last two terms on the right-hand side become a convolution in reciprocal space that couples evanescent waves with different momenta, 
\begin{align}\label{S2_floquet_VIE}
\mathbf{M}(\mathbf{q},z,\omega)
=
\mathbf{M}^{\rm fl}(\mathbf{q},z,\omega)
+
\sum_{\mathbf{G}}
\overleftrightarrow{\chi}_{\mathbf{G}}(z,\omega)
\cdot
\left[
\int dz'\,
\overleftrightarrow{G}_0(\mathbf{q}-\mathbf{G},z,z',\omega)
\cdot
\mathbf{M}(\mathbf{q}-\mathbf{G},z',\omega)
-
\overleftrightarrow{N}\cdot
\mathbf{M}(\mathbf{q}-\mathbf{G},z,\omega)
\right],
\end{align}
where the susceptibility $\overleftrightarrow{\chi}_{\mathbf{G}}(z,\omega)$ is defined by the periodic metasurface geometry,
\begin{align}
\overleftrightarrow{\chi}_{\mathbf{G}}(z,\omega)
=
\frac{1}{A_{\rm uc}}
\int_{uc} d^2\boldsymbol{\rho}\,
\overleftrightarrow{\chi}(\boldsymbol{\rho},z,\omega)
e^{-i\mathbf{G}\cdot\boldsymbol{\rho}},
\qquad
\overleftrightarrow{\chi}(\boldsymbol{\rho},z,\omega)
=
\sum_{\mathbf{G}}
\overleftrightarrow{\chi}_{\mathbf{G}}(z,\omega)
e^{i\mathbf{G}\cdot\boldsymbol{\rho}},
\end{align}
where $A_{\rm uc}$ is the area of the unit cell of metasurfaces in real space. To solve this VIE, we now discretize the CoFeB metasurface with thickness $t$ into $N_z$ cells centered at $z_k=\left(k-\frac{1}{2}\right)\Delta z, \, k=1, \cdots,N_z,$ with $\Delta z = t/N_z$ along the z-direction, and consider $N_G+1$ reciprocal lattice points $\mathbf{G}_i, \, i=0,\cdots,N_G,$ with $\mathbf{G}_0=\mathbf{0}$.  We can rewrite the metasurface VIE, 
\begin{multline}\label{S2_floquet_VIE_2}
\mathbf{M}(\mathbf{q}-\mathbf{G}_i,z_k,\omega)
=
\mathbf{M}^{\rm fl}(\mathbf{q}-\mathbf{G}_i,z_k,\omega)
+
\sum_{j=0}^{N_G}
\overleftrightarrow{\chi}_{\mathbf{G}_j-\mathbf{G}_i}(z_k,\omega)
\cdot \\
\left[
\sum_{l=1}^{N_z}\, \Delta z \, 
\overleftrightarrow{G}_0(\mathbf{q}-\mathbf{G}_j,z_k,z_l,\omega)
\cdot
\mathbf{M}(\mathbf{q}-\mathbf{G}_j,z_l,\omega)
-
\overleftrightarrow{N}\cdot
\mathbf{M}(\mathbf{q}-\mathbf{G}_j,z_k,\omega)
\right],
\end{multline}
where we do not require $\mathbf{q}$ in the first Brillouin zone. This equation is the central VIE in the $(\mathbf{q},z)$ representation for the spin noise metasurfaces. It shows explicitly that, for metasurfaces, the Langevin magnetization noise at given momenta $\mathbf{q}$ does not propagate independently, but is reshaped by the patterned metasurface geometry and gets coupled to evanescent waves with different in-plane momenta $\mathbf{q}-\mathbf{G}_i$. Denoting $\mathbf{M}_{i,k}=\mathbf{M}(\mathbf{q}-\mathbf{G}_i,z_k,\omega)$, $ \mathbf{M}^{\rm fl}_{i,k}=\mathbf{M}^{\rm fl}(\mathbf{q}-\mathbf{G}_i,z_k,\omega)$, $\overleftrightarrow{\chi}_{ij}
=
\overleftrightarrow{\chi}_{\mathbf{G}_j-\mathbf{G}_i}(\omega)$, and $
\overleftrightarrow{G}_{j,kl}
=
\overleftrightarrow{G}_0(\mathbf{q}-\mathbf{G}_j,z_k,z_l,\omega)$, we rewrite the discretized metasurface VIE Eq.~\eqref{S2_floquet_VIE_2} as,
\begin{align}\label{S2_discrete_VIE}
\mathbf{M}_{i,k}
=
\mathbf{M}^{\rm fl}_{i,k}
+
\sum_{j=0}^{N_G}
\overleftrightarrow{\chi}_{ij}
\cdot
\left[
\sum_{l=1}^{N_z}
\Delta z \,
\overleftrightarrow{G}_{j,kl}
\cdot
\mathbf{M}_{j,l}
-
\overleftrightarrow{N}
\cdot
\mathbf{M}_{j,k}
\right].
\end{align}
Stacking all $\mathbf{M}_{i,k}$ and $\mathbf{M}^{\mathrm{fl}}_{i,k}$ into vectors $\mathbb{M}$ and $\mathbb{M}^{\mathrm{fl}}$, we write
\begin{align}\label{AT_definition}
\mathbb{A}(\mathbf{q},\omega)\mathbb{M}
=
\mathbb{M}^{\rm fl},
\qquad
\mathbb{M}
= \mathbb{A}^{-1}(\mathbf{q},\omega) \mathbb{M}^{\rm fl} =
\mathbb{T}(\mathbf{q},\omega)\mathbb{M}^{\rm fl}, \qquad \mathbb{M} = \begin{bmatrix}
    \mathbf{M}_{0,1}\\
    \mathbf{M}_{0,2}\\
    \vdots \\
    \mathbf{M}_{0,N_z}\\
    \mathbf{M}_{1,1}\\
    \mathbf{M}_{1,2}\\
    \vdots \\
    \mathbf{M}_{N_G,N_z}
\end{bmatrix}, \qquad \mathbb{M}^{\rm fl} = \begin{bmatrix}
    \mathbf{M}^{\rm fl}_{0,1}\\
    \mathbf{M}^{\rm fl}_{0,2}\\
    \vdots \\
    \mathbf{M}^{\rm fl}_{0,N_z}\\
    \mathbf{M}^{\rm fl}_{1,1}\\
    \mathbf{M}^{\rm fl}_{1,2}\\
    \vdots \\
    \mathbf{M}^{\rm fl}_{N_G,N_z}
\end{bmatrix},
\end{align}
with
\begin{align}\label{S2_A_matrix}
\mathbb{A}
=
\mathbb{I}
+
\mathbb{X}\mathbb{N}
-
\mathbb{X}\mathbb{G}.
\end{align}
The large matrices $\mathbb{A}, \mathbb{X}, \mathbb{N}, \mathbb{G}$ have the dimension $3N_z(N_G+1) \times 3N_z(N_G+1)$, with $N_z(N_G+1) \times N_z(N_G+1)$ blocks and each block a $3\times 3$ matrix,
\begin{align}
\mathbb{G}_{(i,k),(j,l)}
=
\delta_{ij}
\,\Delta z\,
\overleftrightarrow{G}_0(\mathbf{q}-\mathbf{G}_j,z_k,z_l,\omega),
\qquad
\mathbb{X}_{(i,k),(j,l)}
=
\delta_{kl}
\overleftrightarrow{\chi}_{\mathbf{G}_j-\mathbf{G}_i}(\omega),
\qquad
\mathbb{N}_{(i,k),(j,l)}
=
\delta_{ij}\delta_{kl}\overleftrightarrow{N},
\end{align}
where we treat the pair $(i,k),(j,l)$ as composite block indices following the stack convention in Eq.~(\ref{AT_definition}).

The matrix $\mathbb T(\mathbf q,\omega)=\mathbb A^{-1}(\mathbf q,\omega)$ contains the self-consistent dressing of the Langevin magnetization sources by the patterned metasurface. Physically, it captures the metasurface-defined couplings of highly evanescent waves with different momenta, which reshape their contributions to the dephasing noise spectrum $J_{\rm em}(\omega)$. To obtain $J_{\rm em}$, we now connect the metasurface spin fluctuations to the near-field magnetic dyadic Green functions $\overleftrightarrow{G}_m$. This requires the fluctuation-dissipation theorem (FDT) at two levels, the field FDT relates the magnetic field fluctuations $\langle \mathbf{H}^{\mathrm{fl}}\mathbf{H}^{\mathrm{fl},\dagger}\rangle$ to the imaginary part of the magnetic dyadic Green function $\mathrm{Im} \, \overleftrightarrow{G}_m$, while the material FDT connects the fluctuations of Langevin magnetization sources $\langle \mathbf{M}^{\mathrm{fl}}\mathbf{M}^{\mathrm{fl},\dagger}\rangle$ to the dissipative response of CoFeB metasurface. The field FDT at the position of the spin qubit $\mathbf{r}_{qb}$ is~\cite{buhmann2013dispersion},
\begin{equation}\label{field_FDT}
\left\langle
\mathbf{H}^{\mathrm{fl}}(\mathbf r_{qb},\omega)\,
\mathbf{H}^{\mathrm{fl},\dagger}(\mathbf r_{qb},\omega')
\right\rangle
=\frac{2\pi\hbar k_0^2}{\mu_0}\,\coth{(\frac{\hbar\omega}{2k_BT})}\,\delta(\omega-\omega')\,
\frac{\overleftrightarrow{G}(\mathbf r_{qb},\omega)-\overleftrightarrow{G}^\dagger(\mathbf r_{qb},\omega)}{2i}.
\end{equation}
In the ultra-subwavelength near-field regime, the fluctuating fields $\mathbf{H}^{\mathrm{fl}}$ at the qubit position $\mathbf{r}_{qb}$ are dominated by the magnetic dipolar fields from the magnetization fluctuations at metasurfaces,
\begin{align}\label{magnetic_dip}
\mathbf{H}^{\mathrm{fl}}(\mathbf{q},z_{qb},\omega)
=
\sum_{k=1}^{N_z}
\Delta z \,
\overleftrightarrow{G}_0(\mathbf{q},z_{qb},z_k,\omega)
\cdot
\mathbf{M}_{0,k}.
\end{align}
The material FDT connects the Langevin magnetization fluctuations $\left\langle 
\mathbf M^{\mathrm{fl}}(\mathbf q,z_k)\,\mathbf M^{\mathrm{fl},\dagger}(\mathbf q',z_l)\right\rangle$ to magnetic susceptibility~\cite{buhmann2013dispersion},
\begin{equation}\label{material_FDT}
\left\langle
\mathbf{M}^{\mathrm{fl}}(\mathbf q,z,\omega)\,
\mathbf{M}^{\mathrm{fl},\dagger}(\mathbf q',z',\omega')
\right\rangle
=\sum_{n=0}^{N_G} \, \frac{(2\pi)^3\hbar}{\mu_0}\,\coth{(\frac{\hbar\omega}{2k_BT})}\,\delta(\omega-\omega')\,
\delta^{(2)}(\mathbf q-\mathbf q' - \mathbf{G}_n)\,\delta(z-z')\,
\widetilde{\overleftrightarrow{\chi}}_{\mathbf{G}_n},
\end{equation}
where we take $\widetilde{\overleftrightarrow{\chi}}_{\mathbf{G}_n} = (\overleftrightarrow{\chi}_{\mathbf{G}_n}-(\overleftrightarrow{\chi}_{\mathbf{G}_n})^\dagger)/2i=(\overleftrightarrow{\chi}_{\mathbf{G}_n}-\overleftrightarrow{\chi}_{-\mathbf{G}_n}^\dagger)/2i$, and $\overleftrightarrow{\chi}_{-\mathbf{G}_n}^\dagger$ is the Fourier expansion of metasurface patterned $\overleftrightarrow{\chi}^\dagger$. We use Eq.~\eqref{AT_definition} to connect the Langevin sources to the dressed magnetization fluctuations $\left\langle 
\mathbf M(\mathbf q,z_k)\,\mathbf M^\dagger(\mathbf q',z_l)\right\rangle$ in metasurfaces via the $\mathbb T$ matrix. Combining Eqs.~(\ref{field_FDT},~\ref{material_FDT},~\ref{magnetic_dip}) and the main text Eq.~(2), we can obtain the nanophotonic dephasing noise spectrum near arbitrary spin noise metasurfaces. 

Meanwhile, for the convenience of numerical calculations, we introduce a much simpler form by rewriting the integrand in terms of the large $3N_z(N_G+1) \times 3N_z(N_G+1)$ matrices. 
We first introduce a large $3(N_G+1) \times 3N_z(N_G+1)$ matrix $\mathbb P$ that propagates metasurface magnetization fluctuations to near-field magnetic field fluctuations,
\begin{align}
\mathbf{H}(\mathbf{q}-\mathbf{G}_i,z_{qb},\omega)
=
\sum_{k=1}^{N_z}
\Delta z \,
\overleftrightarrow{G}_0(\mathbf{q}-\mathbf{G}_i,z_{qb},z_k,\omega)
\cdot
\mathbf{M}_{i,k}.
\end{align}
After stacking all channels following the convention in Eq.~(\ref{AT_definition}),
\begin{align}
\mathbb H = \mathbb P \, \mathbb M,\qquad
\mathbb{P}_{i,(j,k)}
=\delta_{ij}
\,\Delta z\,
\overleftrightarrow{G}_0(\mathbf{q}-\mathbf{G}_i,z_{qb},z_k,\omega), \qquad \mathbb P \in \mathbb C^{3(N_G+1) \times 3N_z(N_G+1)}.
\end{align}
Using Eqs.~(\ref{field_FDT},~\ref{material_FDT}), after some algebra, we find
\begin{align}
\begin{aligned}\label{G_m_compact}
\frac{\mathrm{Im}\,[\overleftrightarrow{G}(\mathbf r_{qb},\omega)+\overleftrightarrow{G}^\intercal(\mathbf r_{qb},\omega)]}{2}=\frac{1}{(2\pi)^2 k_0^2} \, \mathrm{Re} \Biggr[\int_{-\infty}^{\infty}d^2\mathbf{q} \,  \, \mathbb{L}_0 \, \mathbb{P} \, \mathbb{T} \, \widetilde{\mathbb{X}} \, \mathbb{T}^\dagger \, \mathbb{P}^\dagger \, \mathbb{R}_{N_G+1}\Biggr],
\end{aligned}
\end{align}
where $\mathbb T = \mathbb T(\mathbf{q},\omega)$ and we define, 
\begin{equation}
    \widetilde{\mathbb X} = \frac{1}{\Delta z} \frac{\mathbb X-\mathbb X^\dagger}{2i},
\end{equation}
and the role of $\mathbb L_0$ and $\mathbb R_{N_G}$ in Eq.~(\ref{G_m_compact}) is purely to select the physically relevant reciprocal space channels contributing to the nanophotonic dephasing noise spectrum, and to avoid double counting since our integral over $\mathbf{q}$ is not limited to the first Brillouin zone, with
\begin{subequations}
\begin{equation}
\mathbb L_0 = [\overleftrightarrow{I}_3\ 0\ 0\ 0\ 0 \cdots \ 0\,]\in \mathbb C^{3\times 3(N_G+1)},
\end{equation}   
\begin{equation}
\mathbb{R}_{N_G+1} = [\,\overleftrightarrow{I}_3\, e^{-i (\mathbf G_0-\mathbf G_{0}) \cdot \boldsymbol{\rho}} \ \ \overleftrightarrow{I}_3\,e^{-i(\mathbf G_1-\mathbf G_{0}) \cdot \boldsymbol{\rho}} \ \ \cdots\ \overleftrightarrow{I}_3\,e^{-i(\mathbf G_{N_G} -\mathbf G_{0}) \cdot \boldsymbol{\rho}}\,]
\in \mathbb C^{3\times 3(N_G+1)},
\end{equation}
\end{subequations}
and $\overleftrightarrow{I}_3$ is the $3 \times 3$ identity matrix. From Eq.~(2) in the main text, the nanophotonic dephasing noise spectrum $J_{em}(\omega)$ near arbitrary spin noise metasurfaces is, 
\begin{align}\label{S2_Jem_Gm}
J_{em}(\mathbf{r}_{qb},\omega)
=
\frac{\mu_0}{\hbar \pi^2}
\coth\left(\frac{\hbar\omega}{2k_BT}\right) \, 
\mathbf{m}\cdot \mathrm{Re}\,\Big[ 
\int_{-\infty}^{\infty}d^2\mathbf{q} \, \mathbb{L}_0 \, \mathbb{P} \, \mathbb{T} \, \widetilde{\mathbb{X}} \, \mathbb{T}^\dagger \, \mathbb{P}^\dagger \, \mathbb{R}_{N_G+1} \Big]
\cdot
\mathbf{m}^{\dagger}.
\end{align}
We note that this Eq.~(\ref{S2_Jem_Gm}) is the local nanophotonic dephasing noise spectrum at position $\mathbf{r}_{qb}$ near arbitrary periodic spin noise metasurfaces. Its structure follows directly from the physical sequence derived above, where $\widetilde{\mathbb X}$ gives the dissipative Langevin spin noise source in the patterned CoFeB, $\mathbb T$ dresses this source through the self-consistent metasurface response, and $\mathbb P$ propagates the dressed magnetization fluctuations to the qubit systems. 

In our experiments, the readout laser spot in our measurements is larger than the metasurface period. Therefore, our measured dephasing noise spectrum corresponds to the unit cell averaged near-field nanophotonic dephasing noise spectrum at the distance $d$ of the quantum system (NV ensemble) from the metasurfaces,
\begin{align}\label{S2_unit_cell_average}
J_{em}(d,\omega)=\frac{1}{A_{uc}}\int_{uc}d^2\boldsymbol{\rho}\,J_{em}(\mathbf{r},\omega),
\end{align}
and this unit cell average keeps only the zero spatial Fourier component in Eq.~(\ref{G_m_compact}) since $\int_{uc}d^2\boldsymbol{\rho}\,e^{i\boldsymbol{\Lambda}\cdot\boldsymbol{\rho}}=0,\,\boldsymbol{\Lambda} \neq \mathbf{0}$. This gives us the spatially averaged nanophotonic dephasing noise spectrum at distance $d$ from metasurfaces,
\begin{equation}\label{Jem_avg}
J_{em}(d,\omega)
=
\frac{\mu_0}{\hbar \pi^2}
\coth\left(\frac{\hbar\omega}{2k_BT}\right) \, 
\mathbf{m}\cdot
\mathrm{Re} \, \Big[ \int_{-\infty}^{\infty}d^2\mathbf{q} \, \mathbb{L}_0 \, \mathbb{P} \,  \mathbb{T} \, \widetilde{\mathbb{X}} \, \mathbb{T}^\dagger \, \mathbb{P}^\dagger \, \mathbb{L}_{0}^\dagger \Big]
\cdot
\mathbf{m}^{\dagger}.
\end{equation}
Equation~(\ref{Jem_avg}) is the spatially averaged nanophotonic dephasing noise spectrum at distance $d$ from arbitrary periodic spin noise metasurfaces. 

Meanwhile, for the thin CoFeB metasurfaces (thickness $\sim 5$nm) used in our experiments, we find that neglecting explicit z-dependence (i.e., taking $N_z=1$) is a reasonable approximation. Under this assumption, we can greatly simplify Eqs.~(\ref{Jem_avg}) and obtain the main text Eq.~(4), 
\begin{align}\label{S2_final_Jem}
J_{\rm em}(d,\omega)
=
\frac{\mu_0}{\hbar\pi^2}
\coth\left(\frac{\hbar\omega}{2k_BT}\right)
\mathbf{m}\cdot
\mathrm{Re}\left[
\int d^2q
\sum_{\mathbf{G},\mathbf{G}'}
\overleftrightarrow{G}_0(\mathbf{q},d)
\cdot
\overleftrightarrow{\mathcal{T}}_{\mathbf{G}}(\mathbf{q})
\cdot
\mathcal{I}\mathit{m}\,[
\overleftrightarrow{\chi}_{\mathbf{G},\mathbf{G}'}(\omega)]
\cdot
\overleftrightarrow{\mathcal{T}}_{\mathbf{G}'}^{\dagger}(\mathbf{q})
\cdot
\overleftrightarrow{G}_0^{\dagger}(\mathbf{q},d)
\right]
\cdot
\mathbf{m}^{\dagger},
\end{align}
where
$\mathcal{I}\mathit{m} \overleftrightarrow{\chi}_{\mathbf{G},\mathbf{G}'} = \frac{1}{\Delta z}\widetilde{\overleftrightarrow{\chi}}_{\mathbf{G}'-\mathbf{G}}$, and 
$\overleftrightarrow{\mathcal{T}}_{\mathbf{G}}(\mathbf{q}) = \mathbb{T}_{(0,1),(\mathbf{G},1)}(\mathbf{q})$ captures the self-consistent metasurface dressing due to couplings between evanescent waves with different momenta. $\overleftrightarrow{G}_0(\mathbf q, d)=\int dz \overleftrightarrow{G}_0(\mathbf q, z_{qb},z,\omega)$ is the averaged dyadic Green function that connects metasurface magnetization fluctuation to magnetic field fluctuations. Equation~\eqref{S2_final_Jem} makes explicit the three physical ingredients in the problem, the dissipative magnetic susceptibility sets the Langevin spin noise, the metasurface geometry dresses this noise, and the free-space Green function propagates the dressed fluctuations at metasurfaces to the quantum systems (NV centers in our experiments) in the near-field. 

In the main text Fig.~2, we employ Eq.~(\ref{Jem_avg}) to calculate $J_{em}$. We take a discretization with $N_z=3$ and $N_G= 25920$ reciprocal points for both metasurfaces. We note that, under our discretization, obtaining $\mathbb T$ by a direct inversion of $\mathbb A$ is numerically difficult. We thus employ an iterative solver to calculate $\mathbb T$ from $\mathbb A$. Other quantities in Eq.~(\ref{Jem_avg}) can be directly obtained from the metasurface/thin film geometry defined in the main text and CoFeB susceptibility discussed in the following subsection. In addition, since the measured signal in our experiments comes from an NV ensemble with four crystallographic
orientations, we further perform the ensemble average discussed in the Methods section for calculating $J_{em}(\omega)$.

\subsection{Magnetic Susceptibility of CoFeB}
To obtain the nanophotonic dephasing noise spectrum $J_{em}(\omega)$, we also need the CoFeB susceptibility $\overleftrightarrow{\chi}(\omega)$. Here, we follow the standard Landau-Lifshitz-Gilbert formula to derive $\overleftrightarrow{\chi}(\omega)$~\cite{cullity2011introduction}. We start from the free energy $F$ of CoFeB that contains the Zeeman energy and anisotropy energy (perpendicular anisotropy and uniaxial in-plane anisotropy). For the CoFeB coordinate system with the uniaxial in-plane anisotropy direction as the x-axis, we have~\cite{cullity2011introduction},
\begin{equation}
    F=\frac{1}{2}\mu_0 M_s^2 \mathcal{M}_z^2 - K_{\bot} \mathcal{M}_z^2 - \mu_0 M_s \boldsymbol{\mathcal{M}} \cdot \mathbf{H}_{\mathrm{ext}} - K_{\parallel} \mathcal{M}_x^2,
\end{equation}
where $\boldsymbol{\mathcal{M}}$ is the normalized magnetization direction vector, $M_s$ is the saturation magnetization, $K_\perp$ is the perpendicular anisotropy constant, and $K_\parallel$ is the in-plane anisotropy constant. The effective field $\mathbf{H}_{\mathrm{eff}}$ is,
\begin{equation}\label{Heff}    \mathbf{H}_{\mathrm{eff}}= -\frac{1}{\mu_0 M_s} \frac{\partial F}{\partial \boldsymbol{\mathcal{M}}}=  \mathbf{H}_{\mathrm{ext}} - (M_s - \frac{2K_{\bot}}{\mu_0 M_s})\mathcal{M}_z\mathbf{\hat{z}} + 2\frac{K_{\parallel}}{\mu_0 M_s}\mathcal{M}_x\mathbf{\hat{x}} =\mathbf{H}_{\mathrm{ext}} - M_{\mathrm{eff}}\mathcal{M}_z\mathbf{\hat{z}} + H_k \mathcal{M}_x\mathbf{\hat{x}}.
\end{equation}
Substituting Eq.~\eqref{Heff} into the Landau-Lifshitz-Gilbert equation, we have~\cite{cullity2011introduction},
\begin{equation}\label{LLG}    \frac{d\boldsymbol{\mathcal{M}}}{dt}=-\mu_0\gamma \boldsymbol{\mathcal{M}} \times \mathbf{H}_{\mathrm{eff}}+\alpha_{\rm loss} \boldsymbol{\mathcal{M}}\times\frac{d\boldsymbol{\mathcal{M}}}{dt},
\end{equation}
where $\alpha_{loss}$ is the Gilbert damping factor and $\gamma$ is the gyromagnetic ratio of CoFeB. 

The susceptibility tensor can  then be conventionally solved
from the linear response theory from Eq.~\eqref{LLG}, by considering a small perturbation of $\delta \boldsymbol{\mathcal{M}}$ under a small external magnetic field $\delta\mathbf{H}_{\rm ext}$~\cite{cullity2011introduction}. We find magnetic resonance frequencies in the absence of external magnetic fields,
\begin{equation}
    \omega_1=\gamma\mu_0 (H_k+M_{\mathrm{eff}}) , \qquad \omega_2=\gamma\mu_0 H_k,
\end{equation}
and the magnetic susceptibility tensor $\overleftrightarrow{\chi}(\omega)$,
\begin{subequations}
\begin{equation}
    \overleftrightarrow{\chi}(\omega) = \mu_0 \begin{bmatrix}
        1 & 0 & 0 \\
        0 & \chi_{yy} & \chi_{yz} \\
        0 & \chi_{zy} & \chi_{zz}           
    \end{bmatrix}.
\end{equation}
\begin{equation}
    \chi_{yy}=M_s\frac{-\gamma \mu_0 (\omega_1-i\alpha_{\rm loss}\omega)}{\omega^2-(\omega_1-i\alpha_{\rm loss}\omega)(\omega_2-i\alpha_{\rm loss}\omega)},
\end{equation}
\begin{equation}
    \chi_{zz}=M_s\frac{-\gamma \mu_0 (\omega_2-i\alpha_{\rm loss}\omega)}{\omega^2-(\omega_1-i\alpha_{\rm loss}\omega)(\omega_2-i\alpha_{\rm loss}\omega)},
\end{equation}
\begin{equation}
    \chi_{yz}=-\chi_{zy}=M_s\frac{i \gamma \mu_0 \omega}{\omega^2-(\omega_1-i\alpha_{\rm loss}\omega)(\omega_2-i\alpha_{\rm loss}\omega)}.
\end{equation}
\end{subequations}
where $\omega_m=\mu_0 \gamma M_s$. In the main text Fig.~2, for fitting experiments, we take the saturation magnetization $\mu_0 M_s=12500\,\mathrm{G}$, effective magnetization $\mu_0 M_{eff}=12500\,\mathrm{G}$, Gilbert damping factor $\alpha=0.019$, and in-plane uniaxial anisotropy field $H_k=5\,\mathrm{G}$, corresponding to values reported in previous works~\cite{thiruvengadam2022anisotropy,tang2017thickness}. The in-plane easy-axis direction $\pi/12$ from the y-axis used in our calculations is determined by fitting to the experimental noise spectra in Fig. 4(e,f). For the momentum-filter plots in Fig. 2(d), we use an in-plane isotropic magnetic response $\overleftrightarrow{\chi} = \mathrm{diag}(\chi_{yy}, \chi_{yy}, \chi_{zz})$ to isolate the geometry-defined momentum filters, excluding contributions from in-plane material magnetic anisotropy.

In addition, at low frequencies ($\sim$MHz)  $\omega \ll \omega_1\, , \omega_2$, we find, $\mathrm{Im}\,\chi_{yy}\sim \omega\,\alpha_{loss}\omega_m/\omega_2^2$, $\mathrm{Im}\;\chi_{zz}\sim \omega\;\alpha_{loss}\omega_m/\omega_1^2$, revealing the scaling  $\mathcal{I}\mathit{m} \overleftrightarrow{\chi} \sim \omega$ considered in the main text.

\subsection{$J_{em}(\omega)$ Near Thin Films and in Macroscopic Cavities}
In this subsection, we provide equations for calculating the $J_{em}(\omega)$ in the main text Fig.~2(a,b).

For thin films, we calculate $J_{em}(\omega)$ from Eq.~\eqref{Jem_avg} by taking the size of the meta-atom to be the same as the unit cell. Meanwhile, we can also consider the dyadic Green function $\overleftrightarrow{G}_m$ calculated from the Fresnel reflection coefficients
$r_{ss}, r_{pp},r_{sp},r_{ps}$~\cite{novotny2012principles},
\begin{align}
\begin{aligned}\label{anacpd}
    \overleftrightarrow{G}_m (\mathbf{r},\omega)=&\frac{i}{8 \pi^2}\int \frac{d \mathbf{q}}{k_z} e^{2i k_z z} \Biggl( \frac{r_{pp}}{q^2}\begin{bmatrix} q_y^2&-q_x q_y&0\\-q_x q_y&q_x^2&0\\0&0&0 \end{bmatrix}+ \frac{r_{ss}}{k_0^2q^2}\begin{bmatrix} -q_x^2 k_z^2 & -q_x q_y k_z^2 & -q_x k_z q^2\\-q_x q_y k_z^2 & -q_y^2 k_z^2 & -q_y k_z q^2\\q^2 q_x k_z & q^2 q_y k_z & q^4 \end{bmatrix} \\
    & \qquad \qquad + \frac{r_{ps}}{k_0 q^2}\begin{bmatrix} q_x q_y k_z & q_y^2 k_z & q_y q^2\\-q_x^2 k_z & -q_y q_x k_z & -q_x q^2\\0 & 0 & 0 \end{bmatrix} + \frac{r_{sp}}{k_0 q^2}\begin{bmatrix} -q_x q_y k_z & q_x^2 k_z & 0\\-q_y^2 k_z & q_y q_x k_z & 0\\ q_y q^2 & - q_x q^2 & 0 \end{bmatrix} \Biggr),
\end{aligned}
\end{align}
where Fresnel reflection coefficients for a thin film can be calculated from the magnetic susceptibility of CoFeB $\overleftrightarrow{\chi}(\omega)$ following Ref.~\cite{khandekar2020new}. We note that Eq.~(\ref{Jem_avg}) and Eq.~(\ref{anacpd}) give us the same results in numerics.

Meanwhile, for Fig.~2(a), we consider an open single-mode macroscopic cavity with cavity
mode frequency $\omega_{cav} = 2\pi \times 0.1\,\mathrm{MHz}$. We take the Green’s function in cavities,
\begin{equation}
    \mathrm{Im}\,\overleftrightarrow{G}_m (\mathbf{r},\omega)=
   \frac{\omega}{6\pi c} \left[ 1+ P \frac{(\omega_{cav}/2Q)^2}{(\omega_{cav}/2Q)^2+(\omega - \omega_{cav})^2} \right] I_3,
\end{equation}
where $\omega/6\pi c$ comes from the vacuum dyadic Green function, and we take cavity Purcell factor $P = 10^5$ and
quality factor $Q = 5 \times 10^5$. Therefore, although microwave resonant cavities are commonly used for Purcell engineering, they cannot provide effective broadband control over low-frequency $J_{em}(\omega)$ necessary for dephasing engineering.

\section{Intrinsic Spin Bath Noise}
In this section, we briefly discuss the intrinsic diamond noise arising from the $V_2$ spin bath in our sample. Following Ref.~\cite{favaro2017tailoring,bar2012suppression}, we model the spin-bath noise as an Ornstein-Uhlenbeck stochastic process, with a Lorentzian noise spectrum  $J_{s-b}(\omega)$,
\begin{equation}
    J_{s-b}(\omega)=\frac{\Delta^2 \tau_c}{\pi}\frac{1}{1+(\omega \tau_c)^2},
\end{equation}
where $\Delta$ is the average coupling between the bath and the NV electron spin and $\tau_c$ is the bath correlation time. From the intrinsic diamond noise measurements in Fig.~4(d), we find $\Delta \approx 4.5\,\mathrm{MHz}, \tau_c \approx 19\,\mathrm{ns}$, comparable to values reported in Ref.~\cite{favaro2017tailoring}, supporting the origin of this intrinsic noise primarily coming from the $V_2$ spin bath. The coupling strength $\Delta$ is determined by the magnetic dipole-dipole interactions between the NV electron spin and the $V_2$ spins. In a photonic environment, this dipolar coupling can roughly follow~\cite{cortes2012quantum},
\begin{equation}
    \Delta \sim \mathrm{Re}\,\overleftrightarrow{G}_m (\mathbf{r}_{V_2}, \mathbf{r}_{NV},\omega) = \mathrm{Re}\,\overleftrightarrow{G}_{refl} (\mathbf{r}_{V_2}, \mathbf{r}_{NV},\omega) + \mathrm{Re}\,\overleftrightarrow{G}_0 (\mathbf{r}_{V_2}, \mathbf{r}_{NV},\omega)
\end{equation}
In the near-field of CoFeB, the distance between NV electron spin and \ce{V2} spins is less than $5\,\mathrm{nm}$~\cite{favaro2017tailoring}, while the distance between the NV spin and metasurface is $\sim45\,\mathrm{nm}$. We thus find  $\mathrm{Re}\,\overleftrightarrow{G}_{refl} (\mathbf{r}_{V_2}, \mathbf{r}_{NV},\omega) \ll \mathrm{Re}\,\overleftrightarrow{G}_0 (\mathbf{r}_{V_2}, \mathbf{r}_{NV},\omega)$. Therefore, the spin noise metasurfaces do not significantly modify the coupling between $V_2$ spin bath and NV centers $\Delta \sim \mathrm{Re}\,\overleftrightarrow{G}_m(\mathbf{r}_{V_2}, \mathbf{r}_{NV})$. In contrast, the bath correlation time $\tau_c$ can be affected by the dissipation of the bath (e.g., relaxation of $V_2$ spins)~\cite{bauch2020decoherence}. Near the spin noise metasurfaces, the relaxation of bath $V_2$ spins is related to   $\mathrm{Im}\,\overleftrightarrow{G}_m(\mathbf{r}_{V_2},\omega_{V_2})$, which can be directly modified by the spin noise metasurfaces. As a result, in the main text, we keep $\Delta'$ close to the intrinsic spin-bath coupling scale, while allowing $\tau_c'$ to vary with photonic environments controlled by spin noise metasurfaces.

\bibliography{supp_reference}